\documentclass[a4paper]{scrartcl}
\usepackage{epsfig}
\usepackage{amssymb}
\usepackage{graphicx}
\usepackage{amsmath}
\usepackage{array}

\usepackage[center]{subfigure}
\usepackage{slashed}
\usepackage{bbm}
\usepackage{mathrsfs}

\newcommand{\Lc}{{\Lambda_c}}
\newcommand{\Lcbar}{{\bar{\Lambda}_c}}
\newcommand{\GeV}{~\mbox{GeV}}
\newcommand{\ba}{\begin{eqnarray}}
\newcommand{\ea}{\end{eqnarray}}
\newcommand{\be}{\begin{equation}}
\newcommand{\ee}{\end{equation}}
\begin{document}

\begin{titlepage}\begin{flushright}
SI-HEP-2011-10
 \vspace{0.6cm}
 \end{flushright}
 \vfill
 \begin{center}
 {\Large\bf
 How much charm can $\overline{P}ANDA$ produce?\\[.5cm]}
{\large\bf
A.~Khodjamirian, Ch.~Klein, Th.~Mannel and Y.-M.~Wang}\\[0.5cm]
{\it  Theoretische Physik 1, Naturwissenschaftlich-Technische Fakult\"at,\\
Universit\"at Siegen, D-57068 Siegen, Germany }
\end{center}
 \vfill
 \begin{abstract}
We consider the production of charmed baryons and mesons in the
proton-antiproton binary reactions  at the energies of the future
$\bar{P}$ANDA experiment. To describe these processes in terms of
hadronic interaction models, one needs strong couplings of the
initial nucleons with the intermediate and final charmed hadrons.
Similar couplings enter the models of binary reactions with
strange hadrons. For both charmed and strange hadrons we employ
the strong couplings and their ratios calculated from QCD
light-cone sum rules. In this method finite masses of $c$ and
$s$ quarks are taken into account. Employing the  Kaidalov's
quark-gluon string model with Regge poles and adjusting  the
normalization of the amplitudes in this model to the calculated
strong couplings, we estimate the production cross section of
charmed hadrons. For $p\bar{p}\to \Lambda_c\bar{\Lambda}_c$ it can
reach several tens of $nb$ at $p_{lab}= 15~\mbox{GeV}$, whereas
the cross sections of $\Sigma_c$ and $D$ pair production are
predicted  to be smaller.

\end{abstract}
\vfill
\end{titlepage}

\section{Introduction}

There is a vivid interest in the cross-section of charmed hadron
production  in the proton-antiproton collisions to be measured by
the future $\bar{P}ANDA$ experiment (see, e.g., \cite{panda}). The
amount of produced charmed mesons and baryons is important for
assessing the ability of this experiment to perform
flavour-physics  oriented studies, such as the measurement of
charm-anticharm mixing, the search for $CP$-violation in $D$
decays or the studies of $\Lambda_c$ decays. A reliable estimate
of the $p\bar{p}\to ~charm $ cross section is however a very
difficult task. The main problem is that the projected energy
range (with the c.m. energy $\sqrt{s}$ varying from 2.25 GeV to
5.47 GeV), being reasonably high for the proton-antiproton
collisions,  is still not far from the threshold of
charm-anticharm production. Several models  of charm production at
these energies can be found in the literature
\cite{KaidV,TitovKampfer,Titov:2011vc,Kroll:1988cd,Goritschnig:2009sq,Braaten,Haidenbauer:2009ad,Haidenbauer:2010nx,Kerbikov:1994xx},
their predictions differing by several orders of magnitude, as
emphasized, e.g. in \cite{Haidenbauer:2009ad}. Especially
difficult is to predict the inclusive charm-anticharm cross
section in the situation where not many exclusive channels are
open. Hence, it is more realistic to assess the exclusive
production of baryons or mesons, such as $p\bar{p}\to \Lc\Lcbar,
\bar{D}D$. A successful model of strange-hadron pair production in
$p\bar{p}$ collisions, which was measured at similar energies, can
serve as a useful tool, provided there is a reliable  way to
replace the model parameters of strange hadrons by the ones for
charmed hadrons. The key parameters in many hadronic models of
these processes are the strong couplings of strange or charmed
baryons with mesons and nucleons. To relate them, the
$SU(4)_{fl}$-symmetry is frequently used in the literature. Note
however, that it is difficult to justify this symmetry in QCD, due
to the large mass difference of the $c$- and $s$- quarks,
$m_c-m_s\gg \Lambda_{QCD}$.

In this paper we employ the strong baryon-meson
couplings of charmed and strange hadrons calculated
from QCD light-cone sum rules (LCSR),
where finite masses of $c$ and $s$ quark are taken
into account. Recently, we calculated \cite{LcDN} the charmed baryon
strong couplings with a charmed meson and a nucleon. In addition
to these results, here we obtain the corresponding strong couplings
of strange hadrons. The nonperturbative inputs in the LCSR method
are the universal nucleon distribution
amplitudes (DA's).
Hence, the extension of our calculation from charmed to
strange hadrons is straightforward and is reduced to a replacement
of the virtual $c$-quark by an $s$-quark in the
underlying correlation functions. In what follows, we also employ  the
ratios of calculated strong couplings which are predicted from LCSR
with smaller uncertainties than the individual couplings.

The results for the strong couplings presented here can be used in
various models of exclusive charm and strange hadron production.
As an application of our results, we use the quark-gluon string
(QGS) model of binary reactions developed by Kaidalov and his
collaborators
\cite{Kaidalov:1981jw,Kaidalov:1981rw,Kaidalov:1999zb,KaidN}. One
version of this model was  already applied  in \cite{KaidV} to
estimate the charm production  cross  section in proton-antiproton
collisions. We refine this model by introducing the helicity
amplitudes and adjusting the two independent strong couplings to
the LCSR estimates.

In what follows, in Sect.~2 we present the LCSR results for strong couplings
of charmed and strange hadrons.
In Sect.~3 we demonstrate how the QGS model works for relatively simple
processes of meson pair production and trace the relation between
the model parameters and strong couplings in QCD.
In Sect.~4 we use the QGS model for $p\bar{p}$ binary reactions
with charmed and strange hadrons, employing the strong couplings from
LCSR and predict the charm production cross sections.
Sect.~5 contains the concluding discussion.
The two appendices contain: (A) the formulae for helicity amplitudes
and (B) the derivation of the absorption factor in $p\bar{p}$ binary processes.

\section{Strong couplings from QCD light-cone sum rules}

The strong couplings  of the $\Lambda_c$-baryon  with
the nucleon and  $D$ or $D^*$ meson  are formally defined as
the following hadronic matrix
elements:
\begin{eqnarray}
\langle \Lc(P-q) |D(-q)  N(P)\rangle &=& g_{\Lc ND} \,
\bar{u}_{\Lambda_c}(P-q)\,i\gamma_5\,u_N(P), \nonumber
\\
\langle  \Lambda_c(P-q) |D^*(-q)  N(P) \rangle &=&
\bar{u}_{\Lambda_c}(P-q)\left(g^V_{ \Lambda_c
ND^*}\slashed{\epsilon}+i\,\frac{g^T_{ \Lambda_c ND^*}}{m_{
\Lambda_c}+m_N}\sigma_{\mu\nu}\epsilon^\mu q^\nu\right)u_N(P). \,
\hspace{0.5 cm} \label{eq:CC1}
\end{eqnarray}
Note that the above couplings are defined in  \cite{LcDN}
as residues at the $D^{(*)}$ and $\Lambda_c$ poles in double
dispersion relations  for the correlation functions with on-shell
nucleon state, hence  all three hadrons are on their mass-shell.
The same definitions are valid for the $\Sigma_c$-baryon couplings
as well as for the corresponding strange hadrons with the
following replacements:  $\Lc(\Sigma_c)\to \Lambda (\Sigma) $
and $D^{(*)}\to K^{(*)}$ in the above.

The $\Lambda_c ND^{(*)}$ and  $\Sigma_c ND^{(*)}$ strong couplings
were calculated from LCSR in \cite{LcDN}, where one can find the
detailed description of the sum rule derivation. The results for
the strong couplings which will be used in this paper are
collected in Table~\ref{tab:res}. Note that in \cite{LcDN} two
different interpolating currents for $\Lambda_c$ and $\Sigma_c$
baryons were used. With the procedure of eliminating the negative
parity baryons suggested in that paper, the results agree within
the uncertainties. In this paper we will only use the strong
couplings  obtained for the pseudoscalar interpolating current for
$\Lambda_{(c)}$ and Ioffe current for $\Sigma_{(c)}$,
respectively, because the sum rules in these cases have a
comparatively lower background of higher states. In
Table~\ref{tab:res}  also the ratios of strong couplings obtained
from LCSR are presented, generally they have smaller estimated
uncertainties, because of the common inputs used in the sum rules.

Turning to strange hadrons, we employ the same LCSR method as in
\cite{Braun:2000kw,Braun:2006hz} and, replacing $c$-quark with the
$s$-quark in the correlation function, calculate the $\Lambda
NK^{(*)}$ and $\Sigma NK^*$ couplings. The inputs used in LCSR
consist of universal nucleon DA's which are taken from
\cite{Lenz:2009ar} and explained in detail in \cite{LcDN}. In
particular we use for the virtual $c$ quark in the correlation
function the value $m_c(m_c)=1.28 \pm 0.03$ GeV. The
flavour-specific input parameters which we adopt here for the sum
rules involving strange hadrons are: the  strange quark mass
$m_s(\mbox{2 GeV})= 98 \pm 16$ MeV and the renormalization scale
$\mu_{s} =1.0 \pm 0.2 ~{\rm GeV}$. Furthermore, one and the same
range $M^2=2.0\pm 0.5$ GeV$^2$ of the Borel parameter in the
$\Sigma$ and $\Lambda$ channels is adopted, whereas for the $K^*$
channel we use $\widetilde{M}^2=1.0\pm 0.5 $ GeV$^2$. The
threshold parameter in the LCSR for $\Lambda$($\Sigma$) strong
couplings is taken $s_0=2.55\pm 0.10$ GeV$^2$ ($s_0=2.75\pm 0.10$
GeV$^2$). The criteria of choosing the input parameters and the
quark-hadron duality ansatz in LCSR are the same as the ones used
and discussed in \cite{LcDN}. The two-point QCD sum rules for the
decay constants of $\Lambda$ and $\Sigma$ baryons with
pseudoscalar and Ioffe currents respectively, are taken from
\cite{Liu:2008yg}. Using the same definitions and notation as for
the decay constants of charmed baryons in \cite{LcDN}), we obtain:
\begin{eqnarray}
\lambda_{\Lambda}^{\mathcal{(P)}}=(0.87^{+0.23}_{-0.13}) \times
10^{-2} \,\, {\rm GeV^2}\,, ~~
\lambda_{\Sigma}^{\mathcal{(I)}}=(2.6^{+0.3}_{-0.2} ) \times
10^{-2} \,\, {\rm GeV^2} \,.
\end{eqnarray}
The resulting estimates  of the strange baryon strong couplings
and their ratios obtained from LCSR are presented in
Table~\ref{tab:res}.
Note that $\Lc$ and $\Sigma_c$ belong to different $SU(3)$ multiplets
(as opposed to $\Lambda$ and $\Sigma$), and this circumstance
explains a substantial difference between the ratios of tensor and vector
strong couplings for these baryons.
\begin{table}[t]
\begin{center}
\begin{tabular}{|c|c||c|c||c|c|}
\hline&&&&&\\[-3mm]
Strong  & LCSR  &Strong  &LCSR&Ratio & LCSR \\
coupling&estimate&coupling&estimate&of couplings&estimate\\[1mm]
(charmed)&&(strange)&&$\big(\frac{\mbox{charmed}}{\mbox{strange}}\big)$&\\
\hline
&&&&&\\[-3mm]
$g_{\Lambda_c  N  D} $ & $10.7^{+5.3}_{-4.3}$ &$g_{\Lambda N K} $&
$7.3^{+2.6}_{-2.8}$&
$\frac{g_{\Lambda_c  N  D}}{g_{\Lambda N K}} $&$1.47^{+0.58}_{-0.44}$\\[3mm]
\hline &&&&&\\[-3mm]
$g^V_{\Lambda_c  N  D^{\ast}} $ &
$-5.8^{+2.1}_{-2.5}$&$g^V_{\Lambda N K^{\ast}} $&$-6.1^{+2.1}_{-2.0}$&$\frac{g^V_{\Lambda_c  N  D^{\ast}}}{g^V_{\Lambda N K^{\ast}}} $
&$0.95^{+0.35}_{-0.28}$\\[3mm]
$g^T_{\Lambda_c  N  D^{\ast}}$ & $3.6^{+2.9}_{-1.8}$& $g^T_{\Lambda N K^{\ast}} $& $12.8^{+5.8}_{-5.2}$&&\\[3mm]
$\frac{g^T_{\Lambda_c  N D^{\ast}}}{g^V_{\Lambda_c  N D^{\ast}}} $&
$-0.63^{+0.16}_{-0.28}$
&$\frac{g^T_{\Lambda  N K^{\ast}}}{g^V_{\Lambda  N K^{\ast}}}$& $-2.1^{+0.5}_{-0.6}$&&\\[3mm]
\hline
&&&&&\\[-3mm]
$g_{\Sigma_c  N  D} $ & $1.3^{+1.0}_{-0.9}$ & $g_{\Sigma  N  K} $
& $1.1^{+0.6}_{-0.5}$ &&  \\[3mm]
\hline &&&&&\\[-3mm]
$g^V_{\Sigma_c  N  D^{\ast}} $ &$1.0^{+1.3}_{-0.6}$ &$g^V_{\Sigma N K^{\ast}} $
&$1.7^{+0.9}_{-0.8}$&$\frac{g^V_{\Sigma_c  N  D^{\ast}}}{g^V_{\Sigma N K^{\ast}}} $&$0.56^{+0.42}_{-0.20}$\\[3mm]
$g^T_{\Sigma_c  N  D^{\ast}} $ & $2.1^{+1.9}_{-1.0}$&$g^T_{\Sigma
N K^{\ast}} $&$3.6^{+1.5}_{-1.2}$ &
&\\[3mm]
$\frac{g^T_{\Sigma_c N D^{\ast}}}{g^V_{\Sigma_c N D^{\ast}}} $ & $2.1\pm 0.5$ &$\frac{g^T_{\Sigma N K^{\ast}}}{g^V_{\Sigma N K^{\ast}}} $ & $2.1 ^{+0.6}_{-0.3}$&&\\[3mm]
\hline
\end{tabular}
\end{center}
\caption{\it Numerical results for the strong couplings of charmed \cite{LcDN}
and strange baryons and their ratios obtained from LCSR with nucleon
DA's.}
\label{tab:res}
\end{table}

In the literature (see e.g., \cite{TitovKampfer}), the strong
couplings of charmed hadrons are estimated assuming $SU(4)_{fl}$
symmetry and equating dimensionless couplings, for example,
assuming $g^{V(T)}_{\Lambda_c  N  D^{\ast}}\simeq
g^{V(T)}_{\Lambda N K^{\ast}} $. Such symmetry relations are
difficult to justify from the point of view of QCD. Indeed,
because of the large mass difference of $c$ and $s$ quarks, the
kinematical factors: masses and four-momenta entering the complete
hadronic  matrix elements  differ significantly. In our approach
we are not relying on any form of the $SU(4)_{fl}$ symmetry.
Still, it is interesting to compare the strong couplings for the
charmed and strange baryons obtained from LCSR and collected in
Table~\ref{tab:res}.  We find that the values of the dimensionless
$g^V$ couplings are in the same ballpark, whereas there is a
significant difference between $g^T_{\Lambda N K^{\ast}} $ and
$g^T_{\Lambda_c N D^{\ast}}$. The strange baryon couplings were
also calculated with the Nijmegen  potential model
\cite{Stoks:1999bz} of low-energy scattering, assuming $SU(3)_{fl}$-symmetry.
Expressed in terms  of the dimensionless  $g$-couplings defined in
(\ref{eq:CC1}) the results of \cite{Stoks:1999bz} with their sign conventions
are:
\begin{eqnarray}
g_{\Lambda N K} = 13.4 \div 17.5 \,,& g^V_{\Lambda  N K^{\ast}} =
-(4.3 \div 6.1)\,,&
g^T_{\Lambda N K^{\ast}} =12.4 \div 16.3, \nonumber \\
g_{\Sigma N K} =-(4.1 \div 5.3)\,, & g^V_{\Sigma  N K^{\ast}} =  -(2.4 \div 3.5)\,,&
g^T_{\Sigma  N K^{\ast}} =  -(1.3 \div 4.6)\,.
\end{eqnarray}
Comparing with our predictions for the strange-baryon
couplings given in Table 1, we observe an agreement for vector-meson couplings
within uncertainties. Also the convention-independent
relative signs of $T$ and $V$ couplings agree.
Meanwhile, the LCSR predictions for
$g_{\Lambda N K}$ and $g_{\Sigma N K}$  are systematically
lower than the intervals for these couplings
obtained in the potential model.

\section{ The QGS model for meson pair production }

In the QGS model, the amplitudes of binary reactions, such as
$p\bar{p}\to \Lambda_c \bar{\Lambda}_c $ or
$p\bar{p}\to \bar{D}D$, are described by planar diagrams
depicted in Fig.~1. These diagrams have a dual interpretation.
From the $s$-channel point of view, annihilation of the slow
$u\bar{u}$ or $d\bar{d}$ pair from the initial proton and
antiproton is followed by a creation of the
$c\bar{c}$-pair. The spectator quarks and antiquarks from
the initial proton and antiproton coalesce with the created quark and
antiquark to form the final state
charmed  hadrons.
\begin{figure}[t]
\centering \hspace{0 cm}
\includegraphics[scale=0.75]{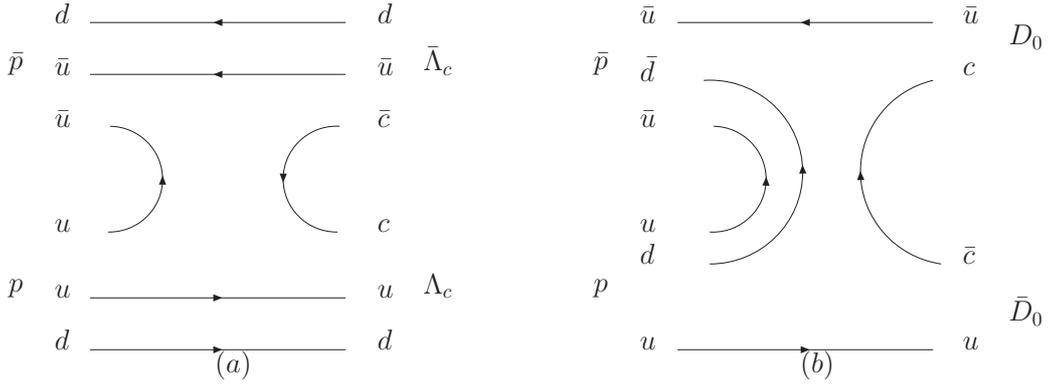}
\caption{\it The planar diagram of charmed baryon
(a) and meson (b) pair production in $p\bar{p}$ collisions.}
\label{fig:planar}
\end{figure}
The intermediate state in $s$-channel represents a sort of a
diquark-antidiquark (Fig.1~a) or quark-antiquark (Fig.1~b) string.
On the other hand, in the $t$-channel a virtual hadronic state
with the quantum numbers of a charmed meson or baryon is
exchanged. In the $s\gg |t|$ limit, this exchange is described by
the dominant Regge pole. For instance, the amplitude of
$p\bar{p}\to \Lambda_c \bar{\Lambda}_c $ is approximated by the
(degenerate)  $D^{*},D^{**}$ Regge-trajectory
$\alpha_{D^*}(t)=\alpha_{D^*}(0)+\alpha'_{D^*}t$ (we use the
linear approximation). The QGS-model parameters are obtained
\cite{KaidV,Kaidalov:1981jw,Kaidalov:1981rw,Kaidalov:1999zb} using
the quark-parton description of the $s$-channel planar diagram.
Replacing the $c$-quark by $s$-quark in the planar diagrams of
Fig.~1 we reproduce the QGS model for the production of strange
baryons and mesons. The strange-hadron pair production cross
section in $p\bar{p}$ collisions calculated in this model
\cite{KaidV} agrees well with the experimental data. Importantly,
there is a strong flavour dependence of the binary  reactions in
QGS model, encoded in the slopes and  intercepts of the Regge
trajectories as well as in the scale factors $s_0$ entering the
Regge amplitudes. The relative suppression of the charmed hadron
production corresponds, in terms of the $s$-channel picture,  to a
comparatively smaller probability to create a heavy
quark-antiquark  pair within the intermediate string.

\begin{figure}[t]
\centering
\includegraphics[scale=0.55]{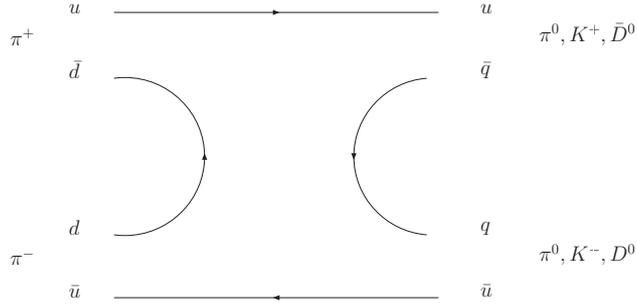}
\caption{\it The planar diagram of $\pi^+\pi^-\to
M \overline{M}$ ($q=u,s,c$ and
$M=\pi^0, K^+, \bar{D}^0$). An additional
diagram with $u\leftrightarrow d$ contributes to the
$\pi^0 \pi^0$-production. } \label{fig:pion}
\end{figure}

To discuss the QGS model in more detail, we first consider a
relatively simple binary reaction involving no spins or
helicities: $\pi^+\pi^-\to M \overline{M}$, with pseudoscalar
mesons ($M=\pi^0,K^+\!,\bar{D}^0$) of various flavours in the
final state and with isospin and/or flavour exchange in
$t$-channel. The planar diagram of this  process is shown in Fig.
\ref{fig:pion}. At large $s$ and small $|t|\ll s $, the scattering
amplitude is written \cite{KaidV} in the following Regge-pole
form: \be T^{(\pi^+\pi^-\to M \overline{M})}(s,t)= g^{(\pi M)}(t)
\frac{s}{\bar{s}}\left(\frac{s}{s_0^{\pi
M}}\right)^{\alpha_{R}(t)-1} , \label{eq:Reggeampl} \ee where
$\bar{s}=1 \GeV^2$ is a universal dimensional factor and the
energy dependence is determined by the Regge trajectory
$\alpha_R(t)$  with the corresponding quantum numbers
($R=\rho(a_2), K^{*(**)},D^{*(**)}$). In the above, $g^{\pi M}(t)$
is the residue function of the momentum transfer squared. In the
QGS model \cite{KaidV} the $\Gamma$-function dependence inspired
by Veneziano duality is adopted: \be
 g^{(\pi M)}(t)=C^{(\pi M)}g_0^2~\Gamma(1-\alpha_R(t))\,.
\label{eq:gt}
\ee
The coefficient $C^{(\pi M)}$ is equal to the number of planar diagrams.

The amplitude (\ref{eq:Reggeampl}) for $\pi^+\pi^-\to
\pi^0\pi^0$ is determined by the $\rho(a_2)$-trajectory:
\be
\alpha_R(t)=\alpha_\rho(t)=\alpha_\rho(0)+\alpha'_\rho t\,.
\label{eq:traject}
\ee
In this case $C^{(\pi\pi)}=2$,
and the numerical values of the intercept $\alpha_\rho(0)$ and slope
$\alpha'_\rho$ taken from \cite{KaidV} are presented in
Table~\ref{tab:reggeM}.

The universal parameter $g_0$ in QGS model
can be related to the $\rho\pi\pi$ strong coupling
defined as:
\be
\langle \pi^-(p_1)\pi^0(p_2)| \rho^-(p_1+p_2)\rangle = g_{\rho\pi\pi}
~\epsilon^{(\rho)}_\mu  (p_2-p_1)^\mu\,,
\label{eq:rhopipi}
\ee
where  $\epsilon^{(\rho)}$ is the polarization 4-vector of $\rho$-meson.
The numerical value $g_{\rho\pi\pi}\simeq 6.0$ with a negligible error
is then  obtained from the measured  \cite{PDG} width:
\be
\Gamma(\rho\to \pi\pi)=\frac{g_{\rho\pi\pi}^2}{6\pi m_{\rho}^2}
(p^*_{\rho\pi\pi})^3\,,
\label{eq:Grhopipi}
\ee
where $p^*_{\rho\pi\pi}=(m_\rho/2)\sqrt{1-4m_\pi^2/m_\rho^2}$ is the 3-momentum of
the pions in the rest frame of the $\rho$.
Combining the product of couplings defined in (\ref{eq:rhopipi})
with the $\rho$ propagator (neglecting the width) we calculate the
$\pi^+\pi^-\to \pi^0\pi^0$ scattering amplitude in a form of a
Feynman diagram with an ``elementary''
$\rho$-meson  exchange in $t$-channel. The result is
\be
T^{(\pi^+\pi^-\to \pi^0\pi^0)}_{diag}(s,t)= g_{\rho\pi\pi}^2
\frac{2s+t-4m_\pi^2}{m_\rho^2-t}\,.
\label{eq:amplFeynm}
\ee
At $s\gg |t|,m_{\pi,\rho}^2$ the above amplitude correctly reproduces
the expected $s^J$ asymptotics, where $J=1$ is the spin of the vector meson
exchanged in $t$-channel. The Regge amplitude (\ref{eq:Reggeampl}),
being analytically continued in
the Mandelstam $\{s,t\}$ plane to $t\sim m_\rho^2$ has to
reproduce the Feynman diagram expression (\ref{eq:amplFeynm})
at $s\gg m_\pi^2,|t|$.
Substituting (\ref{eq:gt}) in  (\ref{eq:Reggeampl}) and expanding
the $\Gamma$-function near $t=m_\rho^2$ where
$\alpha_\rho(m_\rho^2)=1$ we obtain a pole in the variable $t$:
\be
\Gamma(1-\alpha_\rho(t))\simeq
\frac{1}{\alpha'_\rho(m_\rho^2-t)}\,,
\label{eq:gamma} \ee
which
corresponds to the $\rho$-propagator pole in (\ref{eq:amplFeynm}).
Comparing the residues of the amplitudes (\ref{eq:Reggeampl}) and (\ref{eq:amplFeynm}) at large $s$, we obtain
\be
\frac{C^{(\pi
\rho)}g_0^2}{\alpha'_\rho\bar{s}}=2g^2_{\rho\pi\pi}\,.
\label{eq:g2}
\ee
Numerically, the above equation at the values
$\alpha'_\rho=0.9$ and $g^2_0/(4\pi)=2.7$ adopted in \cite{KaidV}
correctly reproduces the experimental value of $g_{\rho\pi\pi}$.
\begin{table}
\begin{center}
\begin{tabular}{|c|c|c|c|c|c|}
\hline process&Regge & intercept & slope
&scale param.&\\
&pole&$\alpha_R(0)$&$\alpha'_R(\mbox{GeV}^{-2})$&$s_0^{\pi M}(\mbox{GeV}^2)$&$C^{(\pi M)}$\\
\hline $\pi^+\pi^-\to \pi^0\pi^0$ &$\rho$&0.46 &0.9 &1.0&2\\ \hline
$\pi^+\pi^-\to K^+ K^-$ &$K^*$&0.32 &0.85&1.25&1\\ \hline
$\pi^+\pi^-\to D^0\bar{D}^0$&$D^*$&$-0.86$ &0.5&3.55&1\\ \hline
\end{tabular}
\end{center}
\caption{\it Parameters of Regge trajectories \cite{KaidV} involved in the
$\pi^+\pi^-\to M \overline{M}$.}
\label{tab:reggeM}
\end{table}

Repeating the same comparison for the binary reaction
$\pi^+\pi^-\to K^+K^-$ with the strangeness-exchange in the
$t$-channel, we replace the $\rho$ trajectory by the $K^*$
trajectory in the Regge-pole amplitude (\ref{eq:Reggeampl}). The
corresponding parameters of GGS model are presented in Table~\ref{tab:reggeM}. Note that the $SU(3)_{fl}$ violation in this model
(i.e. the effect of a heavier $s$-quark) is reflected in the
parameters of Regge trajectory, and also in the flavour-dependent
normalization scale  $s_0^{\pi K}$, introduced in QGS
approach. We then compare the residue function near the pole
at $t=m_{K^*}^2$ with the diagram containing the $K^*$
propagator and the $K^*K\pi$ strong couplings. This diagram
yields an expression similar to
(\ref{eq:amplFeynm}):
\be
T^{(\pi^+\pi^-\to
K^+ K^-)}_{diag}(s,t)= \frac{g_{K^*K\pi}^2}{m_{K^*}^2-t}
\left(2s+t-2(m_\pi^2+m_K^2)+\frac{(m_K^2-m_\pi^2)^2}{m_{K^*}^2}\right)\,,
\label{eq:amplFeynmD}
\ee
with  the same large $s$ asymptotic
behavior. The relation analogous  to (\ref{eq:g2})  yields $
g_{K^*K\pi} =4.5 $ for the $K^{*0}K^+\pi^-$ strong coupling. This
is very close to the value extracted from the $K^*\to K\pi$ width
\cite{PDG}.

Turning to the charmed meson production in the two-pion
collisions, we consider the amplitude (\ref{eq:Reggeampl}) with
the $D^*$ Regge-trajectory: \be T^{(\pi^+\pi^-\to D^0
\bar{D}^0)}(s,t)= g_0^2\,\Gamma(1-\alpha_{D^*}(t))
\frac{s}{\bar{s}}\left(\frac{s}{s_0^{\pi
D}}\right)^{\alpha_{D^*}(t)-1} . \label{eq:ReggepiD} \ee In the
QGS approach, the flavour-dependence of the amplitude is reflected
by the substantial differences between the slope parameters of
$D^*$ and $\rho (K^*)$ trajectories on one hand, and  between the
scale factors $s_0^{\pi D}$ and $s_0^{\pi \pi(\pi K)}$ on the
other hand, as can be seen from Table~\ref{tab:reggeM}. Hence as
we already mentioned, there is no $SU(4)_{fl}$ symmetry in this
model. Still there remains an important question if the universal
value of $g_0^2$ can be used also in the charm production
amplitude.  The $D^*D \pi$  strong coupling is defined as
\footnote{ Note that $2g_{D^*D \pi}$ is equal to the $D^*D \pi$
coupling defined in \cite{BBKR}.} \be \langle \pi^-(p_1)D^0(p_2)|
D^{*-}(p_1+p_2)\rangle = g_{D^*D \pi} ~\epsilon^{(D^*)}_\mu
(p_2-p_1)^\mu\,, \label{eq:DstDpi} \ee and the  ``elementary''
$D^*$-exchange diagram yields the same expression as in
(\ref{eq:amplFeynmD}), where $K\to D$ and  $K^*\to D^*$ have to be
replaced. Continuing the Regge amplitude to $t=m_{D^*}^2$ and
comparing with the large $s$ limit of the $D^*$-exchange
(\ref{eq:amplFeynmD}), we obtain: \be
\frac{g_0^2}{\alpha'_{D^*}\bar{s}}=2g_{D^*D\pi}^2 \,.
\label{eq:Dstdpi} \ee Substituting the slope of the $D^*$
trajectory and using the universal value $g_0^2$ of the QGS model
\cite{KaidV} we obtain\footnote{Our estimate differs from the
smaller value quoted in \cite{KaidN} and based on the same model.
Note that in this earlier paper a larger value of the slope
$\alpha'_{D^*}=0.64$ was used.}:
$g_{D^*D\pi}=5.8\,$.
Interestingly, this value is close to the interval estimated from
QCD LCSR in \cite{BBKR} taking into account the gluon radiative
corrections \cite{KRWY}: $[g_{D^*D\pi}]_{\rm LCSR}= 5.0\pm 1.75$.
The only existing measurement of the total $D^*$ width combined
with the branching fraction yields a larger result: $
[g_{D^*D\pi}]_{\rm exp.}=8.95\pm 0.15 \pm 0.95$
\cite{Ahmed:2001xc}. We conclude that the $D^*$ trajectory slope
adopted in QGS model  is consistent with the LCSR estimates of the
$D^{\ast}D\pi$ strong coupling.
\begin{figure}[h]
\centering
\includegraphics[scale=0.6]{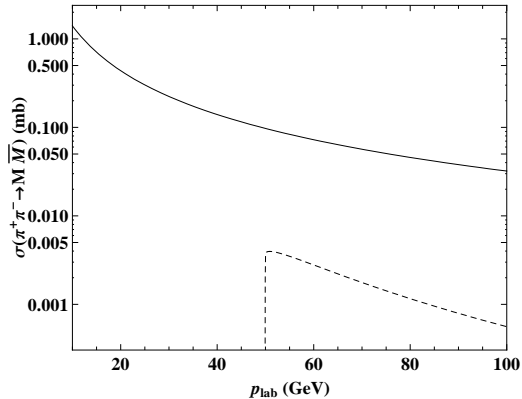}
\vspace{-0.3cm} \caption{ \it Dependence of the cross sections  of
$\pi^+\pi^-\to K^+K^-$ (solid) and
$\pi^+\pi^-\to D^0\bar{D}^0$  (dashed)
on $p_{lab}$ in QGS model.}
\label{fig:compar}
\end{figure}

Note that one of the inputs used to determine the  $D^*$ trajectory of
in QGS model is the Regge trajectory of $J/\psi$,
taken in \cite{KaidV} as:
\be
\alpha_\psi(t)=-2.18 +0.33 t\,,
\label{eq:alphapsi}
\ee
and supported by the QGS model analysis of inclusive charm production.
Here one can mention the estimate of the intercept $\alpha_\psi(0)$
obtained in \cite{AKO}, where  a four-point correlation function of
heavy-quark-currents was first calculated using OPE in terms of
loop diagram and vacuum condensates and then, via optical theorem,
related to the photon-heavy meson scattering
cross section  taken in the Regge form. The comparison of two
representations of the correlation function yields
$\alpha_\psi(0)=-(2 \sim 3)$,
consistent with (\ref{eq:alphapsi}). Note that the
perturbative loop approximation in \cite{AKO} yields $\alpha_\psi(0)=0$,
hence the estimated value of this parameter is entirely determined
by nonperturbative (gluon- and quark-condensate) effects.

To illustrate the QGS model for meson pair production, in
Fig.~\ref{fig:compar}  we present the cross sections of the
processes discussed  in this section. According to our definition
of the scattering amplitude, the differential cross section is:
\be
\frac{d\sigma}{dt}(\pi^+\pi^-\to M \bar{M})= \frac{
|T^{(\pi^+\pi^-\to M \bar{M})}(s,t)|^2}{16\pi s(s-4m_\pi^2)}\,,
\ee
where we substitute the Regge-pole amplitude (\ref{eq:Reggeampl})
for $M=K,D$ and integrate over $|t|$ from the kinematically
allowed minimal $|t_0|$ to $|t_{0}|+\Delta$. Here we choose for
definiteness $\Delta=0.6$ GeV$^2$, so that $|t|$ remains much
smaller than $s$  and hence the Regge-pole description is
applicable. Hereafter we refrain from predicting  total cross
sections, because at large $|t| \sim s$ the behavior of the
scattering amplitudes is governed by mechanisms other than a
simple Regge-pole model. Still, differential cross sections
rapidly decrease with $|t|$, hence, our results for the integrated
cross sections provide an order-of-magnitude estimate also for the
total cross sections. As we can see from Fig.~\ref{fig:compar},
the charmed meson production cross section is strongly suppressed
with respect to the strange-meson one.

\section{ $p\bar{p}$- production of
charmed and strange hadrons }
The QGS model
of the charmed baryon-pair production
in $p\bar{p}$ collision is described by the planar diagram in Fig.~1a.
This amplitude, similar to $\pi\pi\to D\bar{D}$, is approximated
by the $D^*$ Regge-trajectory.
The amplitude of $p\bar{p}\to \Lambda_c\bar{\Lambda}_c$ presented in \cite{KaidV} has a helicity-averaged form
\be
|T^{(p\bar{p}\to \Lambda_c\bar{\Lambda}_c)}(s,t)|=
g^{(p\Lc)}(t)
\frac{s}{\bar{s}}\left(\frac{s}{s_0^{p\Lambda_c}}\right)^{\alpha_{D^*}(t)-1}\,,
\label{eq:Reggeampllambdac}
\ee
where the residue function $g^{(p\Lc)}(t)= C^{p\Lc}g_0^2\Gamma(1-\alpha_{D^*}(t))$
contains the same universal coupling $g_0^2$.  Importantly, the scale
factor $s_0^{p\Lambda_c}$
obtained following \cite{KaidV}
(see Table~\ref{tab:reggeinp}) is not equal to $s_0^{\pi D}$, reflecting
the difference between baryon and meson production in this model.

Here we consider a more elaborated version of the QGS model
for this process with the helicity
amplitudes (see App.~A).
The differential cross section has the following expression
\begin{eqnarray}
 \frac{d\sigma}{d t}(p\bar{p}\to \Lambda_c\bar{\Lambda}_c)= \frac{1}{32 \pi
s(s-4m_p^2)}\bigg
[|H(++,++)|^2+2|H(+-,++)|^2
\nonumber\\
+2|H(++,-+)|^2 +|H(--,++)|^2
+|H(-+,-+)|^2
+|H(+-,-+)|^2 \bigg ] \,, \label{eq:diffLc}
\end{eqnarray}
where in the helicity amplitudes
$H(\lambda_1\lambda_2;\lambda_3\lambda_4)$
the notation $\lambda_{1,2},(\lambda_{3,4})$
denotes the helicities of the proton and antiproton ($\Lc$ and ${\bar{\Lambda}_c}$), respectively.
The $s,t$ dependence of the amplitudes is not shown  for brevity.
At fixed $s$, the region of the momentum transfer
squared $t$ is given by $t_1<t<t_0$, where:
\be
t_{0(1)}=m_p^2+m_{\Lambda_c}^2-\frac{s}{2} +(-) \frac12
\sqrt{(s-4m_p^2)(s-4m_{\Lambda_c}^2)}\,.
\ee
We assume that each helicity amplitude has the Regge form
(\ref{eq:Reggeampllambdac}). The residue functions are
fixed by continuing these Regge
amplitudes to the point $t=m_{D^*}^2$ and matching them to the
helicity amplitudes of $p\bar{p}\to \Lambda_c\bar{\Lambda}_c$ with
an ``elementary'' $D^*$ exchange. The latter are given in App.~A.
(see eq.(\ref{eq:helicityDst})) and contain
two independent strong $\Lambda_c D^* N$
couplings defined in (\ref{eq:CC1}), so that  the coupling
$g_{\Lc ND^*}^V$ ($g_{\Lc ND^*}^T$)
enters  the helicity-nonflip (-flip) amplitudes.
We arrive at the following
expression for the cross section:
\ba
\frac{d\sigma}{dt}(p\bar{p}\to \Lc\bar{\Lambda}_c)&=&
\frac{C_{A}^{(p\bar{p}\to \Lc\bar{\Lambda}_c)}\!(s,t)}{4\pi s(s-4m_p^2)}
|R_{D^*}(s,t)|^2 \nonumber\\
&\times& \Big(|g_{\Lc ND^*}^V|^2
+\frac{|t|}{(m_{\Lc}+m_N)^2}|g_{\Lc ND^*}^T|^2 \Big)^2\,,
\label{eq:dsigdt} \ea where the function \be
R_{D^*}(s,t)=\alpha'_{D^*}\Gamma(1-\alpha_{D^*}(t)) \, s \,
\left(\frac{s}{s_0}\right)^{\alpha_{D^*}(t)-1} \label{eq:dsigmaLc}
\ee is determined by the Regge-pole parameters. As opposed to
$\pi\pi\to M \overline{M}$, where we used the original QGS model
with the universal normalization parameter $g_0^2$, there is now a
more subtle substructure of the Regge amplitudes with the strong
couplings determining the helicity-flip and helicity-nonflip
amplitudes. Furthermore, we modified the above cross section with
respect to (\ref{eq:diffLc}) by multiplying it with the so called
absorption factor $C_{A}^{(p\bar{p}\to \Lc\bar{\Lambda}_c)}$ which
is included following \cite{KaidV}. This factor derived in App.~B
takes into account the initial- and final-state
rescattering of the baryons and antibaryons and suppresses the
cross sections.

The related processes
of charmed baryon production: $p\bar{p}\to \Sigma_c\bar{\Lambda}_c$ and
 $p\bar{p}\to \Sigma_c\bar{\Sigma}_c$,  have a similar
description in QGS model, in particular, they are also
dominated by the same $D^*$ Regge-pole exchange.
Their cross sections  depend on the
combinations of couplings $g^{V,T}_{\Sigma_cND^*}$
and $g_{\Lc N D^*}^{V,T}$. The corresponding
expressions  in terms of helicity amplitudes have
minor differences with respect to (\ref{eq:diffLc})
which we will not discuss here for brevity.
The numerical analysis yields
a substantial suppression of the $\Sigma_c$ production
versus $\Lambda_c$ production, due to the difference in the
strong couplings inferred from LCSR.
This suppression will be discussed
below in more detail.

The charmed-meson production, $ p\bar{p} \to \bar{D} D$, is
described in QGS model by the planar diagram depicted in Fig.~1b.
In this case the $t$-channel exchange involves $\Lambda_c$ and
$\Sigma_c$ Regge-trajectories. Their parameters presented in
Table~\ref{tab:reggeinp} are assumed equal. However according to
our predictions, the strong couplings of these baryons to mesons
and nucleons quite differ from each other, hence there is a
significant difference between the cross sections of charged and
neutral charmed meson-pair production. Indeed, in the planar
diagram model, the process $ p\bar{p} \to D^- D^+$ can only be
mediated by the $\Sigma_c^{++}$ exchange in $t$-channel, whereas
in $ p\bar{p} \to \bar{D}^0 D^0$ both trajectories $\Lambda_c$ and
$\Sigma_c^+$ enter the amplitude. Note that this is a
characteristic feature of the planar diagram mechanism. Inelastic
scattering in the final state ($D^0 \bar{D}^0 \to D^{+} D^{-}$)
due to nonplaner diagrams can enhance $D^{+} D^{-}$ production
cross section. Moreover, in a model where these processes are
mediated by intermediate charmonium states in $s$-channel,
$p\bar{p}\to \{\bar{c}c\} \to D\bar{D}$, there is no correlation
between the flavours of initial and final hadrons, so that both
charged and neutral $D$ mesons are produced with equal rates. Such
a model may indeed work as an additional mechanism for charmed
meson-pair production slightly above the threshold, (see e.g.,
\cite{Kerbikov:1994xx}) but the resulting cross section is
much smaller  than the one generated by $t$-channel exchanges. Let as
also mention that, according to the model \cite{Haidenbauer:2010nx}
based on the baryon-antibaryon potential, the initial state
inelastic interaction could significantly enhance the $D^{+} D^{-}$
production cross section in the near-threshold region.
Therefore, the charged charmed meson cross section
can serve as a useful  check of different charm-production models.

The decomposition in the helicity amplitudes in $ p\bar{p} \to \bar{D}D$
is simpler than for the baryon-pair production because
only the helicities of the initial proton and antiproton are
involved. For the $\bar{D}^0D^0$
production we follow the same method of matching the Regge-pole
amplitude to the ``elementary'' $\Lambda_c$-exchange at $t=m_\Lc^2$
and obtain the cross section:
\begin{eqnarray}
 \frac{d\sigma}{dt}(p\bar{p}\to
 \bar{D}^0 D^0)= \frac{C_{A}^{(p\bar{p}\to\bar{D}^0 D^0 )}\!(s,t)}{32 \pi s(s-4m_p^2)}
|R_{\Lambda_c}(s,t)|^2 (m_{\Lambda_c}^2 -t)|g_{\Lc ND}|^4\,,
\end{eqnarray}
where
\begin{eqnarray}
 R_{\Lambda_c}(s,t)=\alpha'_{\Lambda_c}\Gamma\big({1 \over 2}-\alpha_{\Lambda_c}(t) \big) \, \sqrt{s}
\, \left(\frac{s}{s^{pD}_0}\right)^{\alpha_{\Lambda_c}(t)-1/2},
\end{eqnarray}
and the $\Sigma_c$ exchange contribution is neglected due
to much smaller couplings.
The absorption factor $C_{A}^{(p\bar{p}\to\bar{D}^0 D^0 )}$ in the
above cross section has a form similar to the one in
(\ref{eq:dsigdt}).

\begin{table}
\begin{center}
\begin{tabular}{|l|c|c|c|c|}
\hline & & & & \\[-2mm]
process&Regge pole & $\alpha_R(0)$& $\alpha'_R({\rm GeV}^{-2}$)
&$s_0^{p H}({\rm GeV}^2$)\\
\hline
& & & & \\
$p \bar{p} \to\Lambda_c \bar{\Lambda}_c,  \Sigma_c \bar{\Sigma}_c$ &$D^{\ast}$& $-0.86$ & $0.5$ & $5.76$\\
\hline
& & & & \\[-2mm]
$p \bar{p} \to  \Lambda \bar{\Lambda},\Sigma \bar{\Sigma}$ &$K^{\ast}$& $0.32$ & $0.85$ & $2.43$\\
\hline
& & & & \\[-3mm]
$p \bar{p} \to  D^{0} \bar{D}^{0}$ &$\Lambda_c,\Sigma_c$ &&&\\
& &$-1.82$ &$0.5$ &$3.30$\\
& & & & \\[-4mm]
$p \bar{p} \to  D^{+} D^{-}$ &$\Sigma_c$
&&&\\
\hline
& & & & \\[-3mm]
$p \bar{p} \to  K^{+} K^{-}$  &$\Lambda, \Sigma$
&&&\\[-2mm]
& &$-0.64$ &$0.85$ &$1.93$\\
& & & & \\[-4mm]
$p \bar{p} \to  K^{0} K^{0}$ &$\Sigma$ &&&\\
\hline
\end{tabular}
\end{center}
\caption{\it Parameters of the Regge trajectories determining the $p
\bar{p}$ amplitudes of charmed and strange hadron-pair production
in QGS model \cite{KaidV}.} \label{tab:reggeinp}
\end{table}
\begin{figure}[t]
\centering
\includegraphics[scale=0.6]{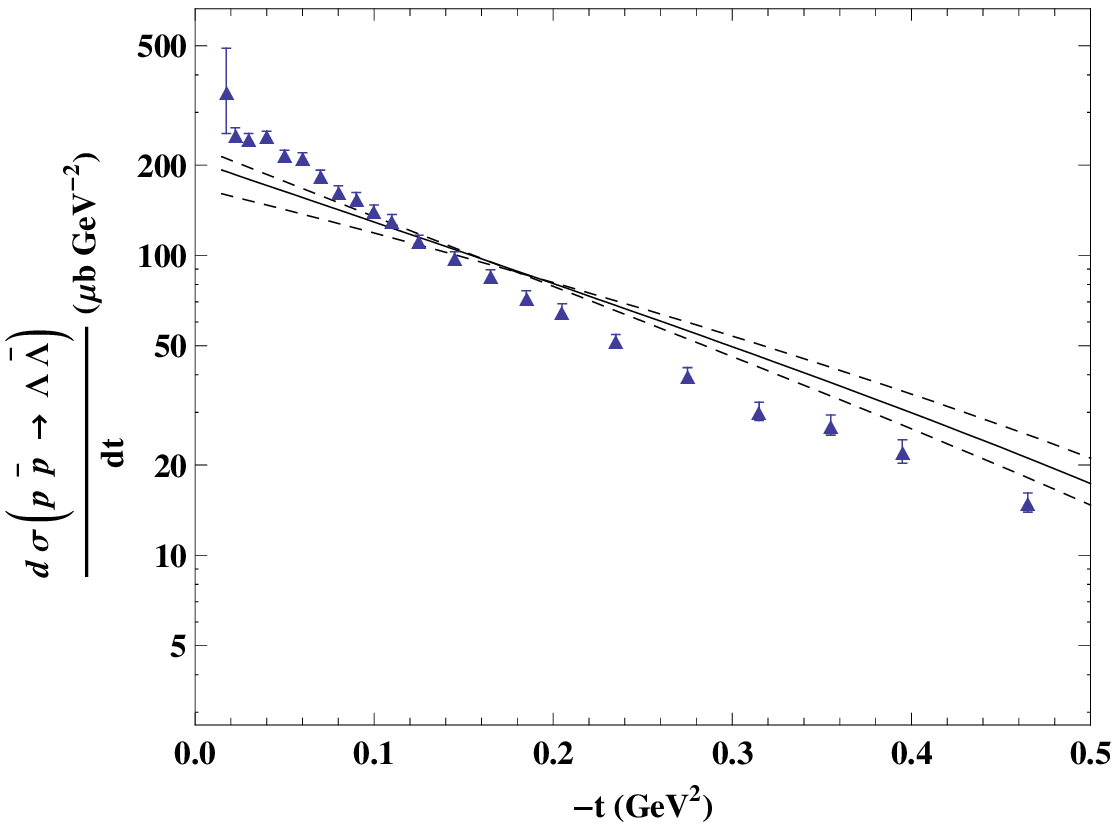}\\\vspace{-0.6cm}
\includegraphics[scale=0.6]{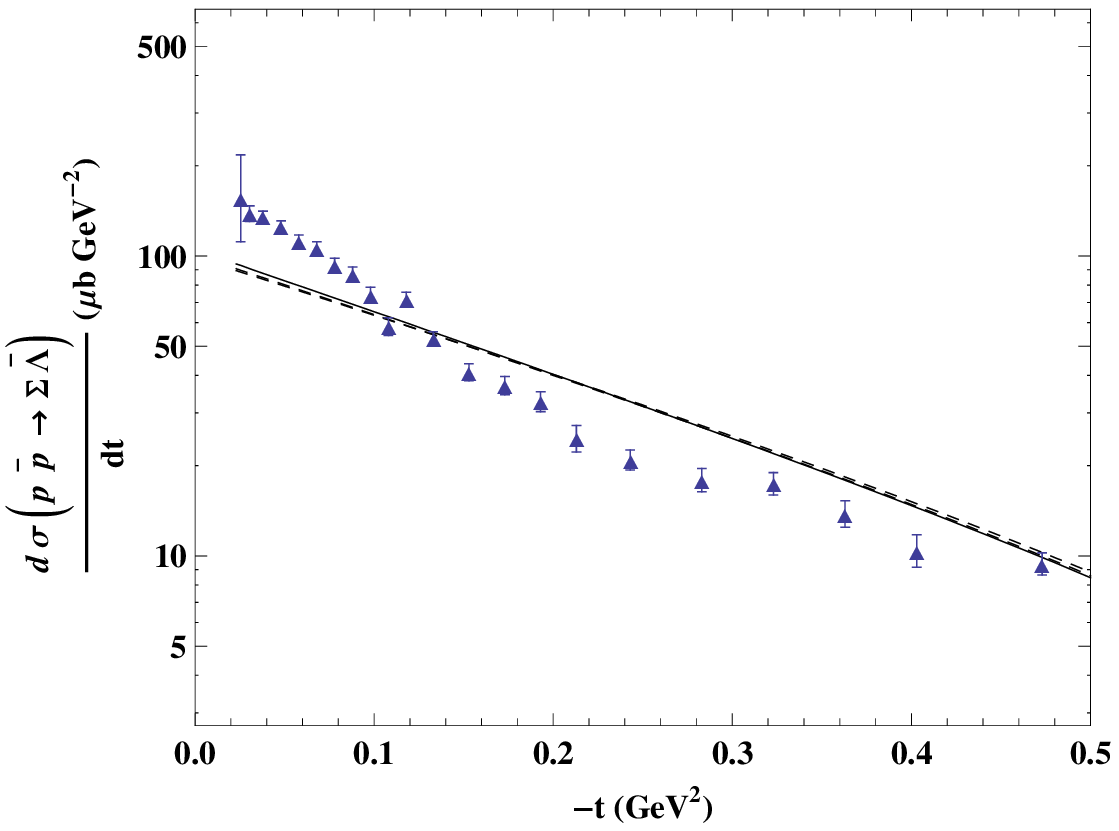}
\includegraphics[scale=0.6]{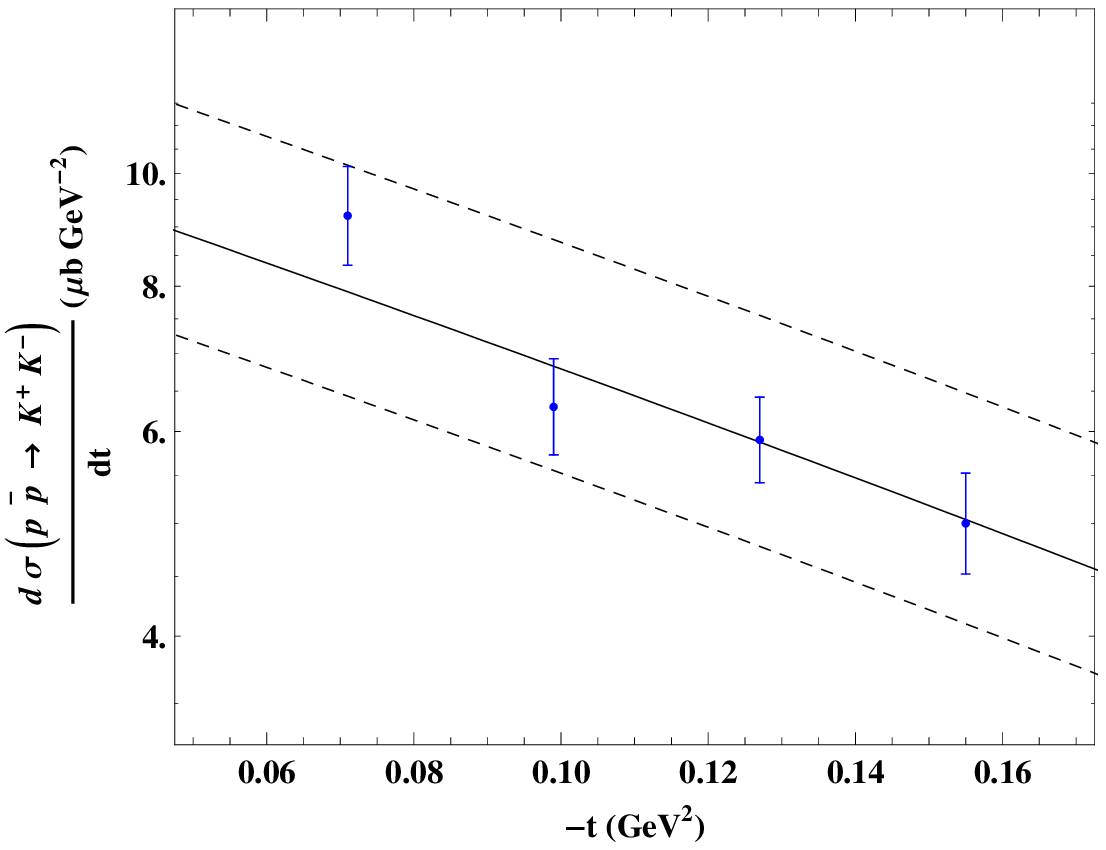}
\vspace{-0.3cm} \caption{\it Differential cross sections  of $ p
\bar{p} \to \Lambda \bar{\Lambda}$ and  $ p \bar{p} \to \Sigma
\bar{\Lambda}$ at $p_{lab}=6 $ GeV, and $ p \bar{p} \to K^{+}
K^{-}$  at $p_{lab}=4 $ GeV. The data points are from
\cite{Becker:1978kk,Brabson:1973qx}. The solid curves
are given by QGS model, with the ratio of tensor to vector strong couplings
from LCSR (dashed curves indicate the uncertainties) and
the vector strong couplings fitted to the
measured cross-section normalization.
}\label{fig:Lamdiff}
\end{figure}

Turning to the numerical analysis of the cross sections
we notice that LCSR predictions for strong couplings have a typical
error of $\sim 50 \%$, hence their fourth powers in
the cross sections introduce large uncertainties.
This mainly concerns the $g^V$-couplings. The ratios $g^T/g^V$
are predicted  from LCSR with much smaller uncertainties and
moreover, the helicity-flip contributions to the cross sections
proportional to $g^T$-couplings are
kinematically suppressed at small $t$.

In order to decrease the uncertainty of the predicted
cross sections for charmed hadrons we consider also
the strange hadron pair-production in $p\bar{p}$ collisions.
Extending the (modified) QGS model to these processes,
allows us to test it, because in this
case some (albeit, quite old) experimental data are available.
Moreover,
we use the fact that the ratios of the strong couplings
of charmed and strange  hadrons given in Table~\ref{tab:res}
have comparatively smaller uncertainties, than the individual
couplings,  because
the same nucleon DA's are used in the sum rules in
both cases  of charmed and strange hadrons.
Hence, we can constrain the couplings of strange hadrons
by fitting the model cross sections to experimental data
and then use the calculated ratios of the couplings
to reproduce the  charm cross sections with  smaller uncertainties.

Let us start with the
process $p\bar{p}\to \Lambda\bar{\Lambda}$. Its cross section
in the QGS model has the
same form as (\ref{eq:dsigdt}),  with the strong couplings
$g_{\Lambda N K^*}^{V,T}$, the $K^*$ Regge-trajectory and
the absorption factor $C_{A}^{(p\bar{p}\to\Lambda\bar{\Lambda})}$.
We first calculate the differential cross section ${d\sigma \over dt}
(p\bar{p} \to \Lambda \bar{\Lambda})$, without taking into account
the absorption factor, and fit the slope of the $t$-dependence to
the exponential form $\exp(-L_{R}(s) |t|)$.
In this cross section we use the ratio of tensor and vector
strong couplings, $g^T_{\Lambda N K^{\ast}}/g^V_{\Lambda N
K^{\ast}}$, obtained from LCSR (see Table~\ref{tab:res}).
The slope $L_{R}$ is then used to
calculate the absorption factor $C_{A}^{(p\bar{p}\to\Lambda\bar{\Lambda})}$
as explained in App.~B.
Note that the overall normalization of the cross
section depending on the vector strong coupling  $g^V_{\Lambda
N K^{\ast}}$ does not play role in this determination.
On the other hand, due to difference of the slopes $L_R$ for
the Regge amplitudes of strange and charmed hadron production
the resulting absorption factor turns out to be almost twice
larger for $p\bar{p}\to \Lambda\bar{\Lambda}$ than for
$p\bar{p}\to \Lambda_c\bar{\Lambda}_c$ at small $t$ (see App.~B),
in accordance with the estimates in \cite{KaidV}.

After inserting the
calculated absorption factor in the differential cross section
of $p\bar{p}\to \Lambda\bar{\Lambda}$, in Fig.\ref{fig:Lamdiff} we
compare the latter with the data points \cite{Becker:1978kk}
at $p_{lab}=6 $ GeV and at small $t$ where we expect the QGS model to work.
Note that not only the  Regge amplitude itself
but also the absorption factor contribute to  $t$-dependence,
making it steeper.  As can be seen from this figure, the agreement
of the shape of the differential cross sections with the data is not very good,
which can possibly be traced back to slightly oversimplified model
of $t$ dependence for this particular (not yet sufficiently large)
energy in our model. Still in the integrated cross section
which is our main interest
here, we expect that the imperfection of the shape does not
play an important role.

As a next step, we fit the overall normalization of the  cross
section to the data and obtain the interval of the vector strong
coupling \be |g^V_{\Lambda N K^{\ast}}|=5.5_{-0.3}^{+0.2},
\label{eq:expgVL} \ee which is within the broader interval of the
LCSR prediction given in Table~{\ref{tab:res}. After that, we
combine the above estimate with the calculated ratio
$g^V_{\Lambda_c N D^{\ast}}/ g^V_{\Lambda N K^{\ast}}$ (see
Table~{\ref{tab:res}) and estimate the vector coupling for the
charm case $|g^V_{\Lambda_c N D^{\ast}}|= 5.2_{-1.6}^{+1.9}$,
again in agreement with the interval of LCSR prediction. We use
the above  ``rescaled'' interval for  $g^V_{\Lambda_c N D^{\ast}}$
in obtaining the charmed baryon cross section (\ref{eq:dsigdt}),
thereby  decreasing the resulting uncertainty. Note that the ratio
$g^T_{\Lambda_c N D^{\ast}}/g^V_{\Lambda_c N D^{\ast}}$ is again
taken from the LCSR prediction. The cross section of $\Lambda_c$
pair production we are interested in is then calculated in two
steps: first we fix the exponential slope $L_R$ in order to obtain
the absorption factor and second, include this  factor in the
cross section. To estimate the cross sections of $ p \bar{p} \to
\Sigma_c \bar{\Lambda}_c,\bar{\Sigma}_c  \Sigma_c$ and $p
\bar{p}\to \bar{D}^0 D^0$, we use a similar procedure employing
the available data on strange hadron pair production (see
Fig.~\ref{fig:Lamdiff}). In particular, fitting of the
corresponding strong couplings yields $|g^V_{\Sigma N
K^{\ast}}|=3.9^{+0.1}_{-0.2}$ and $|g_{\Lambda N K}|=
13.9^{+0.9}_{-0.7}$. The LCSR predictions for these couplings
given in Table~{\ref{tab:res}  are only marginally consistent with
the above intervals. Note that the predictions of the potential
model \cite{Stoks:1999bz} with the scatteing potentials obeying a
(slightly broken) $SU(3)_{fl}$ symmetry, for the same couplings
are  in a better agreement with the fitted values.

Differential cross sections  of $ p \bar{p} \to \Lambda_c
\bar{\Lambda}_c ,\Sigma_c \bar{\Lambda}_c , \Sigma_c
\bar{\Sigma}_c $ and $ p \bar{p} \to D \bar{D}$ are displayed in
Fig. \ref{fig:Lcdiff2} as a function of  $t$ at $p_{lab}=15\, {\rm
GeV}$. As expected, their slope is much smaller than in the case
of strange hadron production. The integrated  cross section $\sigma(t_0,\Delta)$ of
charmed baryon or meson pair-production is defined as the integral
of the differential cross section over the region of small
momentum transfers: $\mbox{max} \{t_1,t_0-\Delta\}<t<t_0$, where
we adopt $\Delta=0.6 \, {\rm GeV^2}$. These cross sections plotted
as a function of $p_{lab}$ in the region accessible to
$\overline{P}ANDA$ are presented in Fig.~\ref{fig:Lctot}. The
summary of our results for the cross sections is also displayed in
Table~{\ref{tab:finres}}.

Let us emphasize that the uncertainties
of the predicted cross sections
are still quite large, even after we narrowed them using the strange hadron
production data. Note that we only quote the uncertainties
stemming from the LCSR estimates of the strong couplings. The QGS model itself
has ``systematical'' uncertainties, which is difficult
to assess quantitatively, as it is the case for any phenomenological
hadronic model  not directly  related to QCD.
The predictive
power of the model concerns mostly  the ratios of cross sections,
where the ``intrinsic'' uncertainties of the method  to a large extent
cancel. One important prediction concerns the suppression
of $\Sigma_c$- with respect to $\Lambda_c$-production cross section.
\begin{figure}[t]
\centering
\includegraphics[scale=0.6]{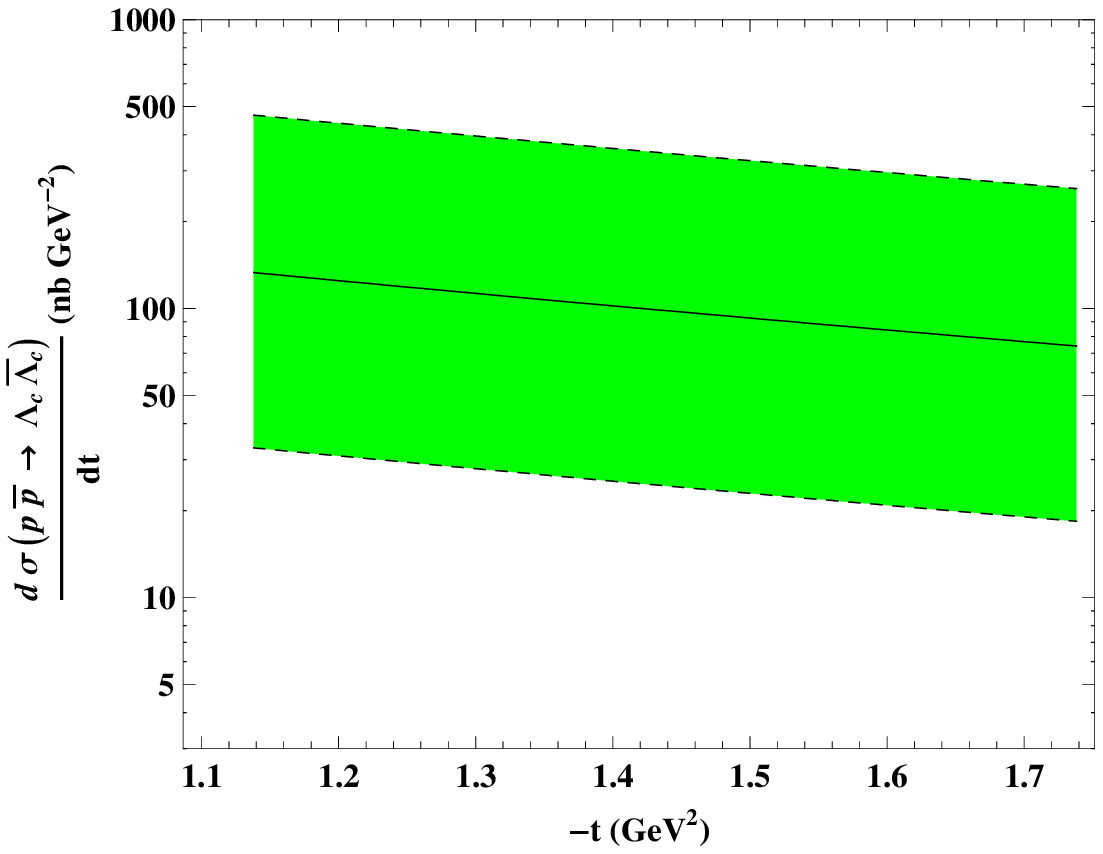}
\includegraphics[scale=0.6]{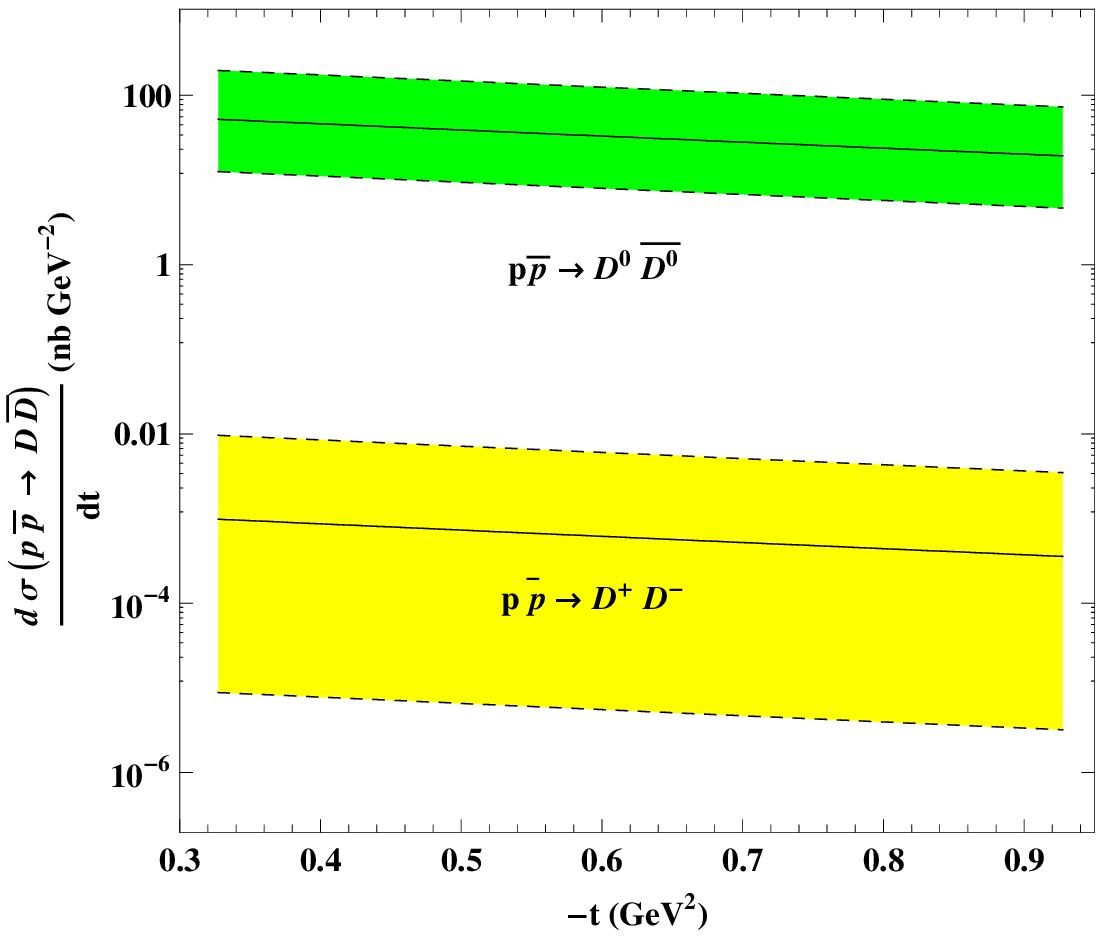}\\
\vspace{-0.3cm} \caption{ \it Differential cross sections  of
$ p\bar{p} \to \Lambda_c \bar{\Lambda}_c $,
and $ p \bar{p} \to D\bar{D}$
at $p_{lab}=15\, {\rm GeV}$ calculated in QGS model. The dashed lines
indicate the uncertainties caused by LCSR estimates of strong couplings.}
\label{fig:Lcdiff2}
\end{figure}

This suppression is more significant than predicted in \cite{KaidV}
where simple relations based on the nonrelativistic quark-diquark model
for these  reactions are used
\begin{eqnarray}
\frac{\sigma(p \bar{p} \to \Lambda_c \bar{\Lambda}_c)}{\sigma(p
\bar{p} \to \Sigma_c \bar{\Lambda}_c)}=
\frac{\sigma(p \bar{p} \to
\Sigma_c \bar{\Lambda}_c)}{\sigma(p \bar{p} \to \Sigma_c
\bar{\Sigma}_c)}= 3 \,.
\label{cross section:relation}
\end{eqnarray}
Our predictions for these ratios
at $p_{lab}= 15 $ GeV are:
\begin{eqnarray}
 \frac{\sigma(p \bar{p} \to \Lambda_c \bar{\Lambda}_c)}{\sigma(p
\bar{p} \to \Sigma_c \bar{\Lambda}_c)}=5.1^{+1.0}_{-2.0}, \, &
\qquad & \frac{\sigma(p \bar{p} \to \Sigma_c
\bar{\Lambda}_c)}{\sigma(p \bar{p} \to \Sigma_c \bar{\Sigma}_c)} =
4.6^{+0.9}_{-1.8} \,.
\end{eqnarray}
\begin{table}
\begin{center}
\begin{tabular}{|c|c|c|}
  \hline
  \hline
    &   &  \\
  channel &  $\frac{ d \sigma}{dt} \big|_{t=t_0} (nb~\mbox{GeV}^{-2})$ & $\sigma(t_0,\Delta) (nb)$ \\
    &   &  \\
  \hline
    &   &  \\
  $ p \bar{p} \to \Lambda_c \bar{\Lambda}_c$ & $ 130 \, (30 \div 470)$ & $60 \,  (15 \div 210)$\\
  &   &  \\
  $ p \bar{p} \to \Sigma_c \bar{\Lambda}_c$ & $24 \,  (5.0 \div 140)$ &  $ 12 \,  (2.0 \div 70)$\\
    &   &  \\
  $ p \bar{p} \to \Sigma_c \bar{\Sigma}_c$ & $ 5.0 \,  (1.0 \div 45)$ &  $ 3.0 \,  (0.4 \div 24)$ \\
    &   &  \\
  $ p \bar{p} \to D^0 \bar{D}^0 $  &  $52 \,  (13 \div 200)$ & $20 \, (5.0 \div 75)$ \\
    &   &  \\
  $ p \bar{p} \to D^+ \bar{D}^- $  &  $< 0.01$ & $<0.01$ \\
    &   &  \\

  \hline
  \hline
\end{tabular}
\end{center}
\caption{\it Differential and integrated cross sections with
$\Delta=0.6~ {\rm GeV^2}$ for charmed hadron production at
$p_{lab} =15 \, {\rm GeV}$.} \label{tab:finres}
\end{table}
Due to the suppression of $\Sigma_c$ couplings
versus $\Lambda_c$ couplings, also the $D^0\bar{D}^0$
production cross section is expected to be significantly larger than
the  $D^+\bar{D}^-$ one. It will be very interesting to test
all these predictions experimentally.

\begin{figure}[t]
\centering
\includegraphics[scale=0.6]{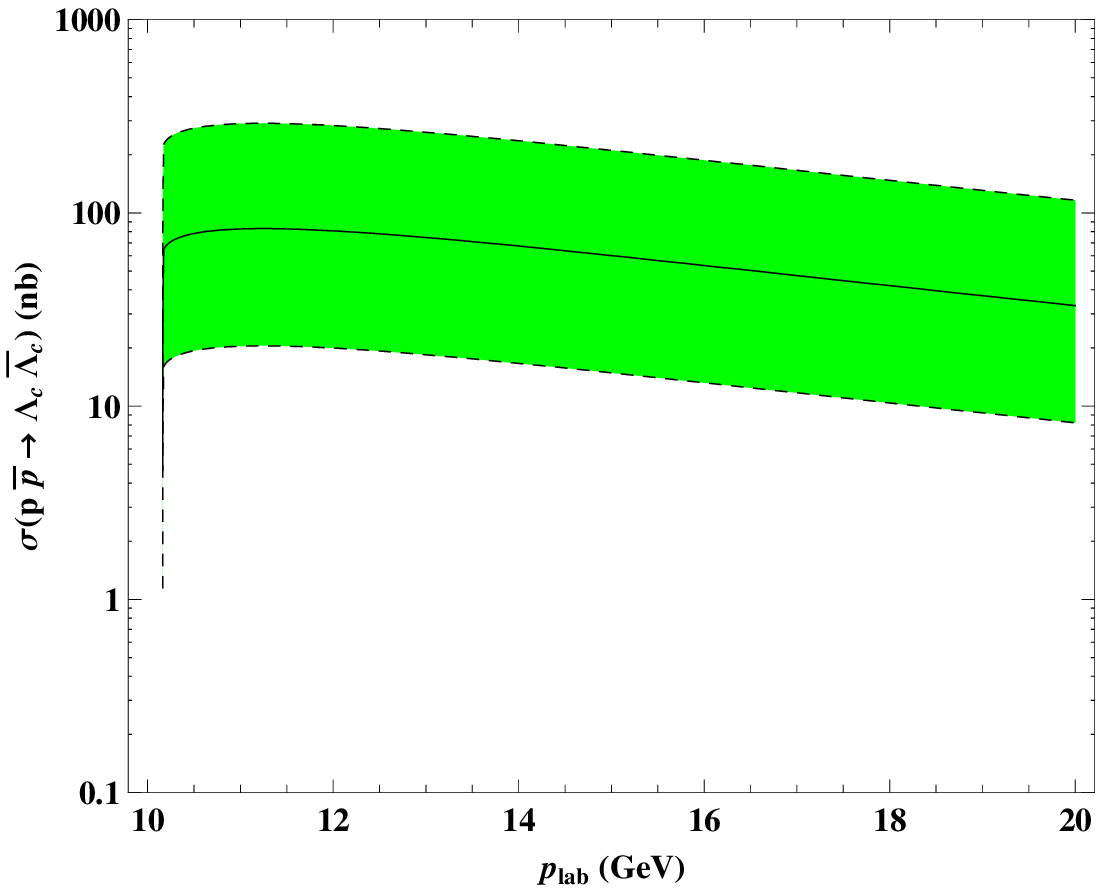}
\includegraphics[scale=0.6]{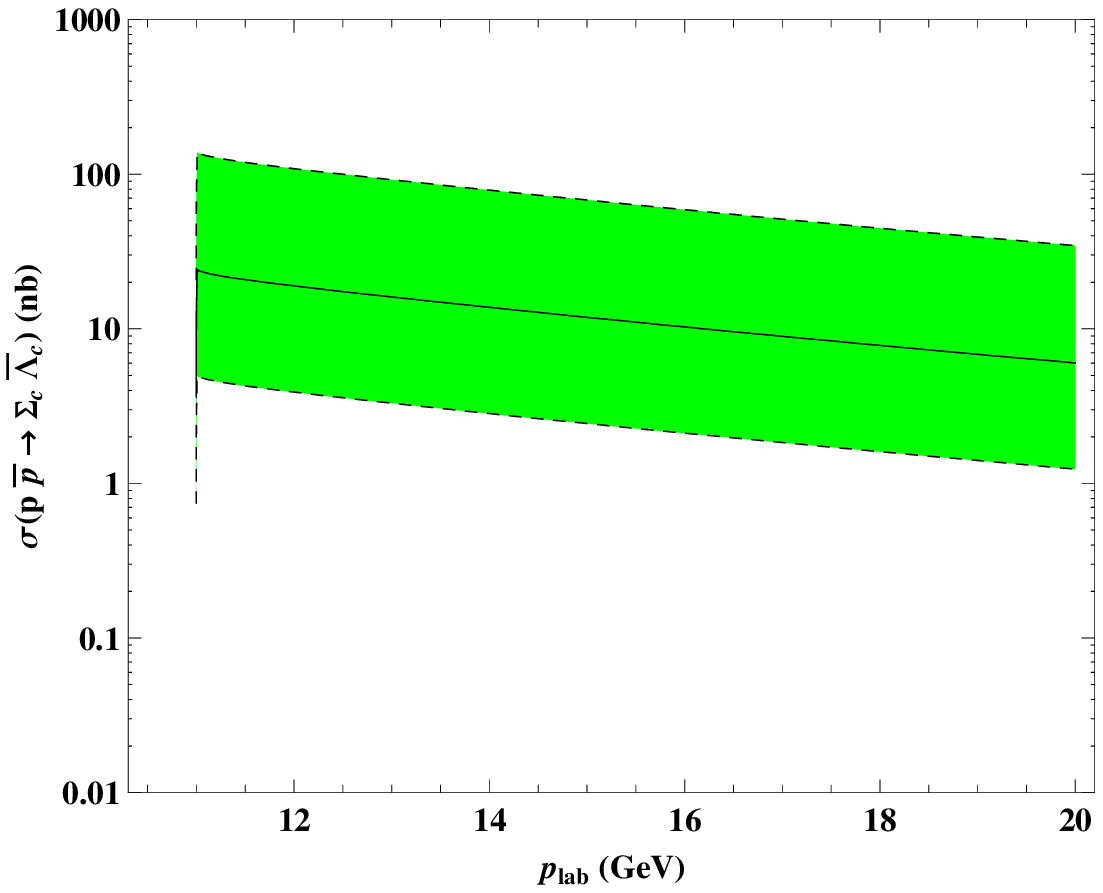}\\
\includegraphics[scale=0.6]{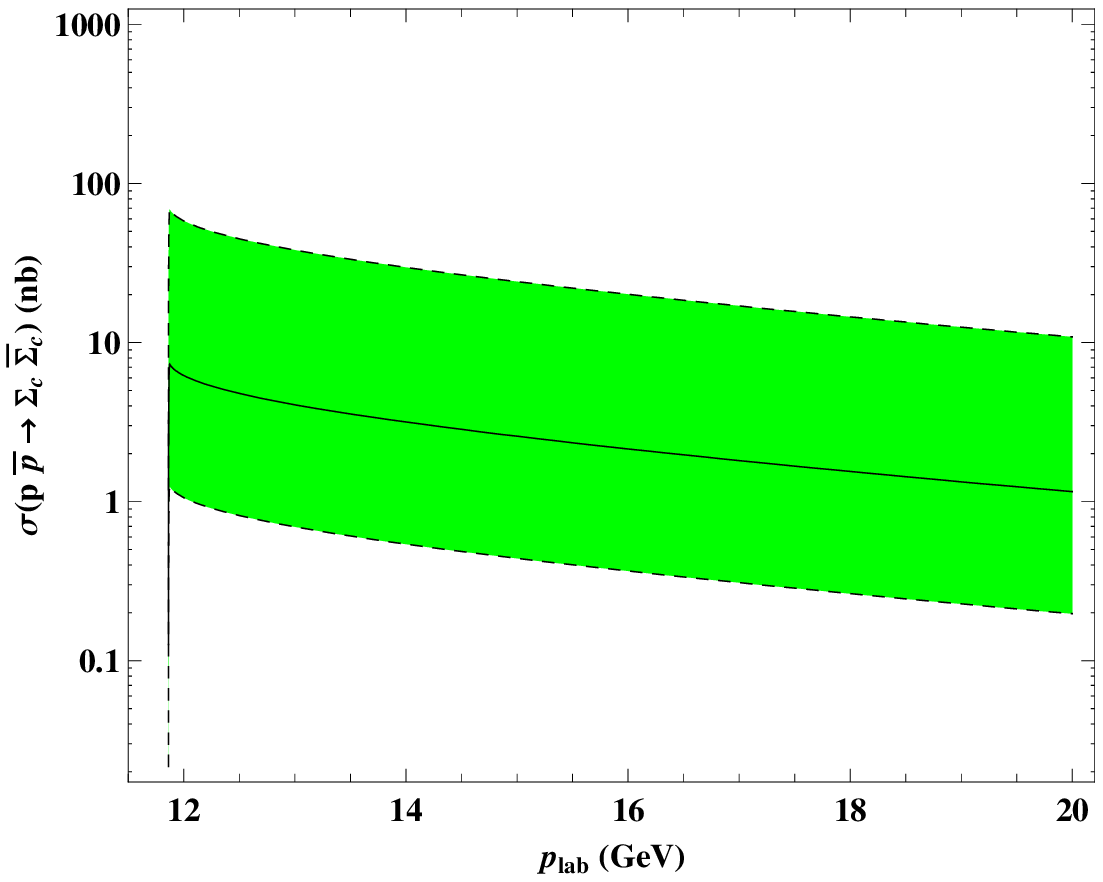}
\includegraphics[scale=0.6]{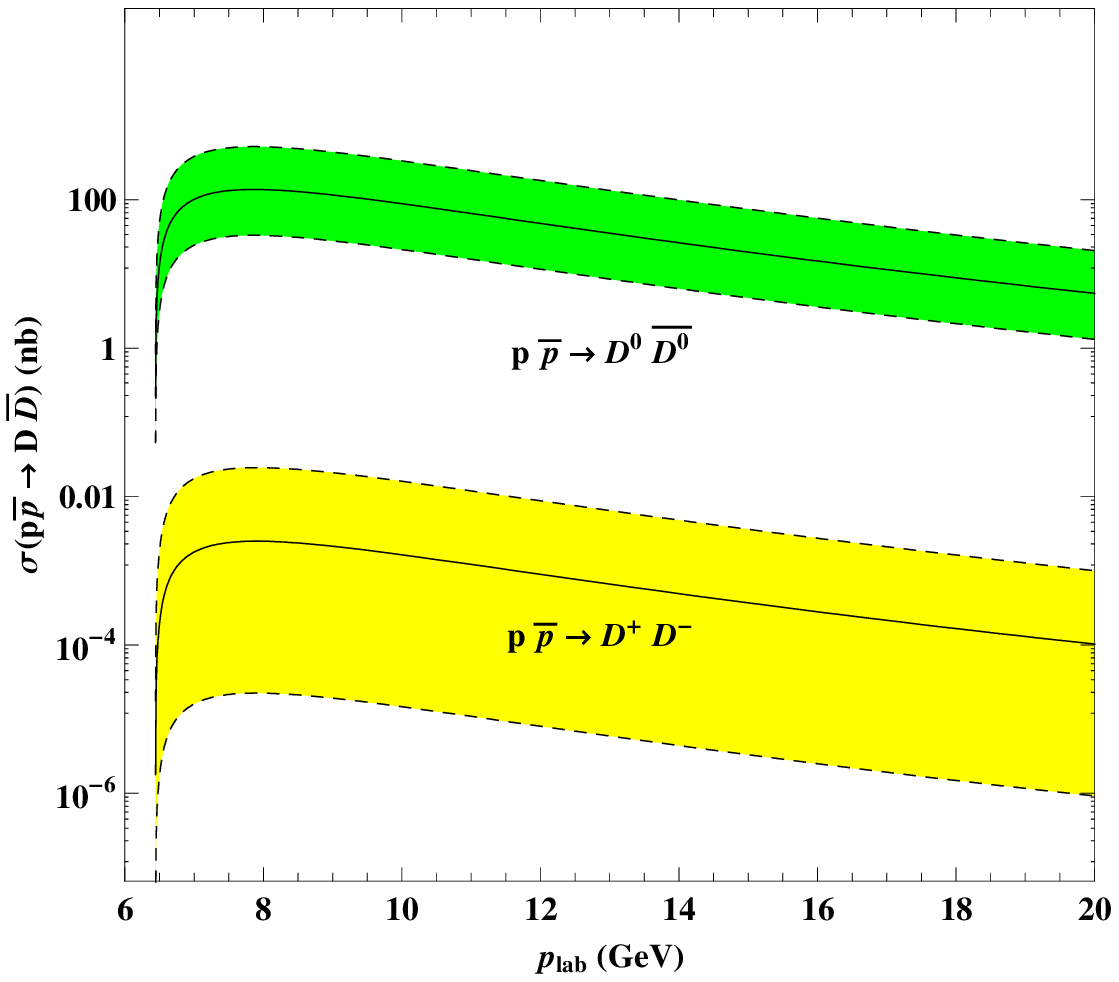}\\
\vspace{-0.3cm} \caption{\it The integrated
cross sections $\sigma(t_0,\Delta)$ of  charmed baryon  and meson pair
production in $ p\bar{p}$ collisions
in QGS model. The dashed lines
indicate the uncertainties introduced by the strong couplings
obtained from LCSR.}
\label{fig:Lctot}
\end{figure}

\section{Conclusion}

In this paper we bring together the strong couplings of charmed and strange
baryons, both predicted within one and the same QCD based method of LCSR.
We have demonstrated that it is possible to avoid
$SU(4)_{fl}$ approximation.
The relations between couplings are nontrivial
because they stem from the nonperturbative dynamics
which is quite different for heavy and light (also strange) quarks.
The LCSR predictions for strong couplings can be significantly improved in future
by calculating radiative gluon corrections and taking into account soft gluon
components of the nucleon DA's.

The main task of this paper is to estimate the charm production
cross sections in $p\bar{p}$ collisions.
For that purpose we have selected the most (in our opinion) ``QCD-friendly''
model
of hadronic reactions, that is, the Kaidalov's QGS model. This
approach has  revealed itself as a very useful tool for
hadronic reactions with different flavours, also for inclusive
production of hadrons. In this paper we  used  a more detailed description
of binary processes with baryons in terms
of helicity amplitudes and employed the strong couplings
of initial protons and final charmed baryons (mesons) with the
intermediate charmed mesons (baryons)
calculated from LCSR. 

Strictly speaking, the QGS model is applicable only at
sufficiently large energies, beyond the upper limit of the
$\bar{P}ANDA$ energy region.  Hence the cross sections calculated
here can only be considered as an order of magnitude estimates,
also because the model is only valid at small momentum transfers
and the absorption factor is only taken in the first
approximation. Still the relations between cross sections are less
influenced by the uncertainties  and are almost independent of the
absorption factors. In future, the model adopted in this paper can
be developed further, taking into account of  the subleading Regge
trajectories and a  more elaborated absorption ansatz.

Finally, turning to the comparison of our results with the
charm-production estimates in the literature, we observe that our
prediction for the dominant $\Lambda_c\bar{\Lambda}_c $ production
is (within estimated uncertainties) consistent with the one
obtained in the original QGS model \cite{KaidV}:  $\sigma
(p\bar{p} \to \Lambda_c\bar{\Lambda}_c) \simeq 100$ nb,
 at $p_{lab}=15~ \mbox{GeV}$, whereas the predictions for  the ratios of
cross sections obtained here and in \cite{KaidV} differ. For example,
we do not exclude a larger charmed meson cross section
than $\sigma (p\bar{p} \to D^0\bar{ D^0}) \simeq 5$ nb predicted
in \cite{KaidV}.

A model of exclusive charm production cross sections
based on the QGS and Regge-poles can be found
in \cite{TitovKampfer}, where the $SU(4)_{fl}$ symmetry is used
and a
different form  of the cross section is adopted,
adding a $t$-dependent dipole ``residual factor''.
Numerically, our predicted intervals
for  the differential cross sections at $t_{0}$  turn out to be larger than
the ones in \cite{TitovKampfer}.

The other models in the literature are based on
radically different approaches. E.g., in
\cite{Haidenbauer:2009ad}}  a hadronic baryon-antibaryon potential
derived  from the coupled channel approach is used, predicting the
cross section of $\Lambda_c\bar{\Lambda}_c$ production up to a few $\mu b$
near the threshold, i.e., much larger than obtained here.
On the opposite side are  the typically smaller cross sections obtained
from perturbative approaches, such as the inclusive charm production
estimate in the parton model \cite{Braaten}
and the approach \cite{Goritschnig:2009sq}
to $p\bar{p} \to \Lambda_c\bar{\Lambda}_c$
employing  distribution amplitudes of initial and final baryons.

Concluding, this paper contains an attempt to apply QCD
predictions for hadronic strong couplings to the models of
exclusive hadronic reactions. Our estimates for charm production
cross sections contain rather large uncertainties. Still even the
lower limit of the cross sections predicted here allows one to
expect an appreciable number of charmed baryons and mesons
produced at $\bar{P}ANDA$. We look forward to other applications
of the strong couplings presented in this paper  and their future
improvements.

\section* {Acknowledgments}
This work is supported by the German Ministry for Education and Research
(BMBF) under contract 06SI9192. T.M. also thanks A. Titov for a discussion
on the subject of the paper.

\section*{Appendix A:  Helicity amplitudes}

The helicity amplitudes
of $\bar{p}p\to \bar{\Lambda}_c\Lambda_c$ scattering
via $D^*$ meson exchange are obtained from the
initial invariant scattering amplitude
\begin{eqnarray}
H(\lambda_1 \lambda_2; \lambda_3 \lambda_4) &=& {1 \over
m_{D^{\ast}}^2 - t} \bar{u}_{\Lambda_c} (p_3, \lambda_3) \bigg[ g^V_{
\Lambda_c ND^*}\slashed{\epsilon}+i\,\frac{g^T_{ \Lambda_c
ND^*}}{m_{ \Lambda_c}+m_N}\sigma_{\mu\nu}\epsilon^\mu q^\nu \bigg]
u_N(p_1,\lambda_1) \nonumber \\
&&  \bar{v}_{\bar{N}}(p_2,\lambda_2) \bigg[ g^V_{ \Lambda_c
ND^*}\slashed{\epsilon}^{\ast}+i\,\frac{g^T_{ \Lambda_c ND^*}}{m_{
\Lambda_c}+m_N}\sigma_{\rho \tau}{\epsilon^{\ast}}^\rho q^\tau
\bigg] v_{\Lambda_c} (p_4, \lambda_4)\,,
\end{eqnarray}
where the baryon bisponors are distinguished by their indices
so that ($p_1, \lambda_1$), ($p_2, \lambda_2$), ($p_3, \lambda_3$)
and ($p_4, \lambda_4$) are the four-momenta and helicities of the
proton, antiproton, $\Lambda_c$ and  $\bar{\Lambda}_c$
respectively; $\epsilon_{\mu}$ is the polarization vector of
the virtual $D^{\ast}$ meson. Generally, there are 16 different
helicity amplitudes for  $\bar{p}p\to \bar{\Lambda}_c\Lambda_c$
process, only six of them are independent due to symmetries
\cite{Bourrely:1980mr}.

In the  c.m. frame of proton-antiproton pair we choose the ${x,z}$ plane
for the process, and the 3-momentum
of proton in the $z$ direction, so that the 3-momentum of $\Lambda_c$
has the angular coordinates $(\theta, \varphi=0)$.
Then the kinematics is as follows:
\begin{eqnarray}
p_1 = {1 \over 2}(  \sqrt{s} , 0 ,0 ,  \sqrt{s- 4 m_N^2}  ) \,,
~~p_2 = {1 \over 2}(  \sqrt{s} , 0 ,0 , -  \sqrt{s- 4 m_N^2}  )
\,,
\nonumber \\
p_3 = {1 \over 2} ( \sqrt{s} ,  \sqrt{s- 4 m_{\Lambda_c}^2} \sin
\theta, 0, \sqrt{s- 4 m_{\Lambda_c}^2} \cos \theta   ) \,,
\nonumber \\
p_4 = {1 \over 2} ( \sqrt{s} ,  -\sqrt{s- 4 m_{\Lambda_c}^2}
\sin \theta,0, -\sqrt{s- 4 m_{\Lambda_c}^2} \cos \theta   )
\end{eqnarray}

Summing over the polarization of $D^{\ast}$ meson:
\begin{eqnarray}
\sum_{\lambda=1, 2, 3, 4} \epsilon^{\mu}(q, \lambda) \epsilon^{\nu
\ast}(q, \lambda) =-g^{\mu \nu} + {q^{\mu} q^{\nu} \over m_{D^*}^2
}\,,
\end{eqnarray}
and substituting explicitly the bispinors with various helicities
in the chosen frame we obtain rather bulky expressions of helicity
amplitudes, which however greatly simplify in the limit of large
$s$ where we compare them with the Regge amplitudes. The six
helicity amplitudes are:
\begin{eqnarray}
H(++,++) &=& { s \over t- m_{D^{\ast}}^2} 2 (g^V_{ \Lambda_c
ND^*})^2 \,, \nonumber \\
H(+-,++) &=& { s \over t- m_{D^{\ast}}^2} {2 \sqrt{-t} g^V_{
\Lambda_c ND^*} g^T_{ \Lambda_cND^*}  \over  m_{ \Lambda_c}+m_N}
\nonumber \\
H(++,-+) &=& -{ s \over t- m_{D^{\ast}}^2} {2 \sqrt{-t} g^V_{
\Lambda_c ND^*} g^T_{ \Lambda_cND^*}  \over  m_{ \Lambda_c}+m_N}
\nonumber \\
H(--,++) &=& -{ s \over t- m_{D^{\ast}}^2} {2 t (g^T_{ \Lambda_c
ND^*})^2   \over (m_{\Lambda_c}+m_N )^2  }
\,, \nonumber \\
H(-+,-+) &=& { s \over t- m_{D^{\ast}}^2} 2 (g^V_{ \Lambda_c
ND^*})^2
\,, \nonumber \\
H(+-,-+) &=& { s \over t- m_{D^{\ast}}^2} {2 t (g^T_{ \Lambda_c
ND^*})^2 \over (m_{ \Lambda_c}+m_N )^2}\,.
\label{eq:helicityDst}
\end{eqnarray}
It is clear that in the large $s$ limit only three helicity
amplitudes are independent. The amplitudes can be related to
each other through the following relations:
\begin{eqnarray}
H(-\lambda_1 -\lambda_2; -\lambda_3 -\lambda_4) &=&
(-1)^{\lambda_1-\lambda_2-\lambda_3+\lambda_4} H(\lambda_1
\lambda_2; \lambda_3 \lambda_4)\,, \nonumber \\
H(\lambda_2 \lambda_1; \lambda_4 \lambda_3)&=&
(-1)^{\lambda_1-\lambda_2-\lambda_3+\lambda_4} H(\lambda_1
\lambda_2; \lambda_3 \lambda_4)\,,
\end{eqnarray}
following from the parity and charge-conjugation
invariance.

\section*{Appendix B:  Absorption factor }

Here we derive the absorption factor $C_{A}(s,t)$, multiplying
the cross section.
The absorption is generated by the (quasi) elastic rescattering  of
the initial proton and antiproton as well as of the final hadron pair,
both are approximated by the pomeron exchange \cite{KaidV}.
Here we make a simplifying assumption that the  elastic rescattering
amplitudes dominate,  they do not change  the helicities
and are the same in the initial
and final states, independent of the flavour
content of the latter.

Consider a process $p\bar{p}\to B\bar{B}$ with generic $B$ hadrons
in the final-state. The amplitude in QGS model, having the form
(\ref{eq:Reggeampllambdac}) has predominantly
exponential behavior at small $t$, the main source of it
is  the Regge-pole factor $(s/s_0)^{\alpha{(t)}}$.
Therefore, the $p\bar{p}\to B\bar{B}$
amplitude can be cast in  the exponential form
\be
T_R(s,t)=f_R(s) \exp\left(-\frac{L_R}{2}|t|\right)\,.
\label{eq:Texp}
\ee
We then switch to the impact parameter representation,
where the 2-dimensional vector $\vec{b}$ is conjugate to the
transverse momentum transfer $\vec{q}_\perp$:
\be
T_R(s,b)=\int
T_R(s,t) \, {\rm
exp}(i\vec{q}_\perp\cdot\vec{b})\frac{d\vec{q}_\perp}{2\pi}\,,
\label{eq:Tb}
\ee
and at high energies $t\equiv q^2\simeq -|\vec{q}|^2$. The
angular integration yields:
\be
T_R(s,b)=\frac12
\int\limits_0^\infty d|t|J_0\left
(\sqrt{|t|}b\right)T_R(s,-|t|)\,,
\ee
where $J_0$ is the Bessel
function and $b=|\vec{b}|$. Substituting the  exponential
representation (\ref{eq:Texp}) in the above integral, and
integrating over $t$ we obtain
\be
T_R(s,b)=\frac{f_R(s)}{L_R}\exp\left(-\frac{b^2}{2L_R}\right)\,.
\label{eq:Trb}
\ee

The rescattering in the initial and final state is dominated  by the
pomeron amplitude $T_P(s,t)$ which is
predominantly imaginary and has an exponential form in the
momentum transfer $T_P(s,t)=T_P(s,0)\exp(-L_P|t|/2)$. The
forward-scattering amplitude is expressed via total $p\bar{p}$ cross section
using  the optical theorem: $\mbox{Im}
T_P(s,0)=2p^*\sqrt{s}\sigma_{p\bar{p}}^{tot}(s)$, where $p^*$ is
the 3-momentum in the c.m. system of the $p\bar{p}$ collision.
Hence one obtains for the pomeron-mediated elastic rescattering
amplitude:
\be
T_{P}(s,t)=2ip^*\sqrt{s}\,\sigma_{p\bar{p}}^{tot}(s)\exp(-L_P|t|/2).
\ee
Employing the same Fourier-transformation to the impact
parameter space as in (\ref{eq:Tb}), it is easy to get the impact parameter
representation for this amplitude: \be
T_{P}(s,b)=\frac{2ip^*\sqrt{s}\sigma_{p\bar{p}}^{tot}(s)}{L_P}
\exp\left (-\frac{b^2}{2 L_P}\right).
\ee

The absorption contribution added to the initial Reggeon amplitude
in the $b$ space yields:
\be
T(s,b)=T_R(s,b)\left[1+i\frac{T_P(s,b)}{8\pi p^*\sqrt{s}} \right]=
T_R(s,b)\bigg[1-\chi(s,b)\bigg]\,, \label{eq:absb}
\ee
where
\be
\chi(s,b)=\frac{\sigma_{p\bar{p}}^{tot}(s)}{4\pi L_P}\exp\left(-\frac{b^2}{2L_P}\right)\,,
\label{eq:chifact}
\ee
and the normalization factor multiplying $T_P$
corresponds to the convention of impact parameter representation
adopted in \cite{KaidV}.
Substituting (\ref{eq:Trb}) in (\ref{eq:absb}) and performing the inverse
Fourier transformation to the $t$-dependent amplitude we finally obtain:
\be
T(s,t)=T_R(s,t)\left[1-
\frac{\sigma_{p\bar{p}}^{tot}(s)}{4\pi(L_P+L_R)}
\exp\left(\frac{L_R^2|t|}{2(L_P+L_R)}\right) \right]\,.
\label{absorbtion:first}
\ee
This expression takes into account absorption in the amplitude
in the first approximation.
To obtain $C_A(s,t)$ one simply has to square the expression in brackets
multiplying $T_R(s,t)$ in the above.
At high energies this effect should be resummed (exponentiated),
however at small $t$'s we are considering here
the resummation effects, as  we checked numerically, are small,
so that the first approximation for $C_A$ is sufficient.

For the numerical evaluation of the absorption factor in
(\ref{absorbtion:first}), the data on $\sigma_{p\bar{p}}^{tot}(s)$
in a parameterized form are taken from \cite{PDG}, so that
$\sigma_{p \bar{p}}^{tot}(s)$ changes from 52.5~mb to 47.9~mb in
the interval $p_{lab}= 10$ GeV to 20 GeV. The slope of the pomeron
mediated elastic $\bar{p}p$ scattering is taken from
\cite{Okorokov:2008zf}, e.g., $L_P=12.1 \,\,{\rm GeV}^{-2}$ at
$p_{lab}= 15$ GeV. Finally, the slopes of Regge-pole amplitudes
fitted to the exponential form (\ref{eq:Texp}) are (in units
GeV$^{-2}$): $L_R=2.6$ ($p_{lab}= 6$ GeV) and $2.5$ ($p_{lab}= 4$
GeV) for $p\bar{p}\to \Lambda \bar{\Lambda}$,
$\Sigma\bar{\Lambda}$ and $K^+K^-$, respectively. For
$p\bar{p}\to\Lambda_c \bar{\Lambda}_c$, $\Sigma_c
\bar{\Lambda}_c$, $\Sigma_c \bar{\Sigma}_c$, $D \bar{D}$, the
corresponding slopes are $L_R=0.6$, $ 0.4$, $0.2$, $1.2$
($p_{lab}= 15$ GeV), respectively. For  numerical illustration, we
quote the absorption factors calculated at the same energy
$p_{lab}=15 $ GeV and at $t=t_0$ for strange and charmed baryon
production: $C_A^{(p\bar{p}\to \Lambda \bar{\Lambda})}=0.09$ and
 $C_A^{(p\bar{p}\to\Lambda_c
\bar{\Lambda}_c)}=0.04$. These values are in agreement with ($t$-averaged)
values presented in \cite{KaidV} and indicate a strong absorption effect
on one hand and a large difference between this effect for strange and
charmed baryons.


\begin{thebibliography}{100}

\bibitem{panda}
  U.~Wiedner,
  Prog.\ Part.\ Nucl.\ Phys.\  {\bf 66} (2011) 477.


\bibitem{KaidV}
  A.~B.~Kaidalov and P.~E.~Volkovitsky,
  Z.\ Phys.\  C {\bf 63} (1994) 517.


\bibitem{TitovKampfer}
  A.~I.~Titov and B.~Kampfer,
  Phys.\ Rev.\  C {\bf 78}, 025201 (2008);



\bibitem{Titov:2011vc}
  A.~I.~Titov and B.~Kampfer,
  arXiv:1105.3847 [hep-ph].



\bibitem{Kroll:1988cd}
  P.~Kroll, B.~Quadder and W.~Schweiger,
  Nucl.\ Phys.\  B {\bf 316} (1989) 373.

\bibitem{Goritschnig:2009sq}
  A.~T.~Goritschnig, P.~Kroll and W.~Schweiger,
  Eur.\ Phys.\ J.\  A {\bf 42} (2009) 43.

\bibitem{Braaten}
  E.~Braaten and P.~Artoisenet,
  Phys.\ Rev.\  D {\bf 79}, 114005 (2009).


\bibitem{Haidenbauer:2009ad}
  J.~Haidenbauer and G.~Krein,
  Phys.\ Lett.\  B {\bf 687}, 314 (2010).

\bibitem{Haidenbauer:2010nx}
  J.~Haidenbauer and G.~Krein,
  Few Body Syst.\  {\bf 50} (2011) 183\,.

\bibitem{Kerbikov:1994xx}
  B.~Kerbikov and D.~Kharzeev,
  Phys.\ Rev.\  D {\bf 51} (1995) 6103.



\bibitem{LcDN}
  A.~Khodjamirian, C.~Klein, T.~Mannel and Y.~M.~Wang,
 JHEP  {\bf 09} (2011) 106.





\bibitem{Kaidalov:1981jw}
  A.~B.~Kaidalov and P.~E.~Volkovitsky,
  Sov.\ J.\ Nucl.\ Phys.\  {\bf 35} (1982) 909
  [Yad.\ Fiz.\  {\bf 35} (1982) 1556];



\bibitem{Kaidalov:1981rw}
  A.~B.~Kaidalov and P.~E.~Volkovitsky,
  Sov.\ J.\ Nucl.\ Phys.\  {\bf 35} (1982) 720
  [Yad.\ Fiz.\  {\bf 35} (1982) 1231];



\bibitem{Kaidalov:1999zb}
  A.~B.~Kaidalov,
  Surveys High Energ.\ Phys.\  {\bf 13} (1999) 265.


\bibitem{KaidN}
  A.~B.~Kaidalov and A.~V.~Nogteva,
  Sov.\ J.\ Nucl.\ Phys.\  {\bf 47} (1988) 321
  [Yad.\ Fiz.\  {\bf 47} (1988) 505].





\bibitem{Braun:2000kw}
  V.~Braun, R.~J.~Fries, N.~Mahnke and E.~Stein,
  Nucl.\ Phys.\  B {\bf 589} (2000) 381
  [Erratum-ibid.\  B {\bf 607} (2001) 433];



\bibitem{Braun:2006hz}
  V.~M.~Braun, A.~Lenz and M.~Wittmann,
  Phys.\ Rev.\  D {\bf 73} (2006) 094019;


\bibitem{Lenz:2009ar}
  A.~Lenz, M.~Gockeler, T.~Kaltenbrunner and N.~Warkentin,
  Phys.\ Rev.\  D {\bf 79}, 093007 (2009).


\bibitem{Liu:2008yg}
  Y.~L.~Liu and M.~Q.~Huang,
  Nucl.\ Phys.\  A {\bf 821} (2009) 80.


\bibitem{Stoks:1999bz}
 V.~G.~J.~Stoks and T.~A.~Rijken,
  Phys.\ Rev.\  C {\bf 59}, 3009 (1999).



\bibitem{PDG}
  K.~Nakamura {\it et al.}  [Particle Data Group],
  J.\ Phys.\ G {\bf 37} (2010) 075021.

\bibitem{BBKR}
  V.~M.~Belyaev, V.~M.~Braun, A.~Khodjamirian and R.~R\"uckl,
  Phys.\ Rev.\  D {\bf 51} (1995) 6177.





\bibitem{KRWY}
  A.~Khodjamirian, R.~R\"uckl, S.~Weinzierl and O.~I.~Yakovlev,
  Phys.\ Lett.\  B {\bf 457}  (1999) 245.


\bibitem{Ahmed:2001xc}
  S.~Ahmed {\it et al.}  [CLEO Collaboration],
  Phys.\ Rev.\ Lett.\  {\bf 87}  (2001) 251801.


\bibitem{AKO}
  A.~Khodjamirian and A.~G.~Oganesian,
  Phys.\ Atom.\ Nucl.\  {\bf 56} (1993) 1720
  [Yad.\ Fiz.\  {\bf 56} (1993) 172].



\bibitem{Becker:1978kk}
  H.~Becker {\it et al.}  [CERN-Munich Collaboration],
  Nucl.\ Phys.\  B {\bf 141} (1978)  48.



\bibitem{Brabson:1973qx}
  A.~Brabson {\it et al.},
  Phys.\ Lett.\  B {\bf 42}  (1972) 287.




\bibitem{Bourrely:1980mr}
  C.~Bourrely, J.~Soffer and E.~Leader,
  Phys.\ Rept.\  {\bf 59} (1980) 95.



\bibitem{Okorokov:2008zf}
  V.~A.~Okorokov,
  arXiv:0811.3849 [hep-ph].


\end{thebibliography}
\end{document}